\documentclass[sigconf]{acmart}

\AtBeginDocument{%
  }

\acmConference[SIGMOD]{Proc. ACM Manag. Data}
\acmISBN{}

\usepackage{algorithmic}
\usepackage{graphicx}
\usepackage{textcomp}
\usepackage{xcolor}
\usepackage{tabularx}
\usepackage{multirow}
\usepackage{subcaption}
\usepackage{makecell} 
\usepackage[ruled, vlined] {algorithm2e}
\usepackage{float}  

\newcommand{\eat}[1]{}
\newcommand{\stitle}[1]{\vspace{1ex}\noindent{\bf #1}}

\newcommand{\eg}{\textit{e.g.}~}

\newtheorem{defn}{Definition}

\usepackage{soul}
\usepackage{xcolor}

\newcommand{\hlred}[1]{#1}
\newcommand{\hlyellow}[1]{#1}
\newcommand{\hlgreen}[1]{#1}
\newcommand{\hlblue}[1]{#1}

\begin{document}

\title{SSCard: Substring Cardinality Estimation using Suffix Tree-Guided Learned FM-Index}


\author{Yirui Zhan}
\affiliation{%
  \institution{Peking University}
  \city{Beijing}
  \country{China}}
\email{zhanyirui@stu.pku.edu.cn}

\author{Wen Nie}
\affiliation{%
  \institution{Huawei Company}
  \city{Beijing}
  \country{China}}
\email{niewen2@huawei.com}

\author{Jun Gao}
\affiliation{%
  \institution{Peking University}
  \city{Beijing}
  \country{China}}
\email{gaojun@pku.edu.cn}

\renewcommand{\shortauthors}{Trovato et al.}

\begin{abstract}
Accurate cardinality estimation of substring queries, which are commonly expressed using the SQL \texttt{LIKE} predicate, is crucial for query optimization in database systems. 
While both rule-based methods and machine learning-based methods have been developed to optimize various aspects of cardinality estimation, their absence of error bounds may result in substantial estimation errors, leading to suboptimal execution plans. In this paper, we propose SSCard, a novel \textbf{S}ub\textbf{S}tring \textbf{Card}inality estimator that leverages a space-efficient FM-Index into flexible database applications. SSCard first extends the FM-Index to support multiple strings naturally, and then organizes the FM-index using a pruned suffix tree. The suffix tree structure enables precise cardinality estimation for short patterns and achieves high compression via a pushup operation, especially on a large alphabet with skewed character distributions. Furthermore, SSCard incorporates a spline interpolation method with an error bound to balance space usage and estimation accuracy. Additional innovations include a bidirectional estimation algorithm and incremental update strategies. Extensive experimental results in five real-life datasets show that SSCard outperforms both traditional methods and recent learning-based methods, which achieves an average reduction of 20\% in the average q-error, 80\% in the maximum q-error, and 50\% in the construction time, compared with second-best approaches.

\end{abstract}




\received{20 February 2007}
\received[revised]{12 March 2009}
\received[accepted]{5 June 2009}

\maketitle

\section{Introduction}

An effective query optimizer is a key factor in the success of a database management system (DBMS), and accurate cardinality estimation results are fundamental for the optimizer to select the efficient plan. 
While substantial progress has been made in cardinality estimation~\cite{lipton1990practical,bruno2004conditional,bruno2001stholes,kiefer2017estimating,tzoumas2011lightweight}, existing research has primarily focused on numerical data, leaving textual data relatively understudied. In this paper, we focus on estimating the \textit{ substring cardinalities}, which are widely used for queries in textual data. 
For example, the predicate \texttt{conference LIKE \%SIGMOD\%} in a SQL query returns all rows with the column ``\textit{conference}‘’ that contains ``\textit{SIGMOD}'' as a substring, regardless of the year or place where the conference is held. 
This statement is widely used in real-life scenarios or query benchmarks, as in the Join Order Benchmark
~\cite{goodplan2015}, there are 81 out of 113 queries containing \texttt{LIKE} statements.

Traditional methods choose suffix tree~\cite{krishnan1996estimating,jagadish1999substring}, $k$-gram~\cite{lee2009approximate} or histograms~\cite{aytimur2021using} as summary data structures to support substring cardinality estimation. While achieving precise results for short query strings, they often depend on statistical assumptions when processing long strings. For example, pruned suffix tree methods~\cite{krishnan1996estimating, jagadish1999substring} assume that the probabilities of strings divided from the long query string are independent, and the cardinality of the long string is computed as the total number of rows multiplied by the product of these probabilities. Things are similar to the $k$-gram-based methods~\cite{lee2009approximate}. For strings longer than $k$, estimations are performed on the combined results of divided strings assuming individual distributions. Although these approaches are easy to interpret, accuracy drops significantly when the assumptions deviate significantly from the actual distribution of substring patterns.

With the rise of deep learning for database systems~\cite{marcus2021bao,chen2023loger,xu2025bqschednonintrusiveschedulerbatch}, learning-based methods are used for cardinality estimation in string queries~\cite{shetiya2020astrid, wang2020monotonic, wang2021face, kwon2022cardinality, aytimur2021using, aytimur2024lplm}. They can be roughly divided into two categories, encoding the underlying data string and encoding the pattern string. The former approaches attempt to capture the distribution of substrings in embeddings, from which the cardinality can be further learned~\cite{shetiya2020astrid}. The latter approaches, including the most recent work~\cite{aytimur2024lplm}, tend to directly predict the cardinality of patterns using the sequence model~\cite{kwon2022cardinality,wang2021face,wang2020monotonic, aytimur2021using, aytimur2024lplm}. Learning-based methods are better suited for handling large volumes of data, as they can benefit from additional training data without requiring an increase in model size. 
However, they cannot provide error bounds for the cardinality estimation. Their performance may also be severely affected when the model is not fully trained due to the limited samples or the vast combination space with a large alphabet.

The string index to support substring occurrences has been extensively studied in other domains~\cite{karkkainen1996lempel, burrows1994block,ferragina2000opportunistic,loukides2023text}, especially in bioinformatics~\cite{delcher1999alignment, delcher2002fast}. Among these, FM-Index~\cite{ferragina2000opportunistic} is an interesting structure which should receive more attention in the database field. An FM-Index contains the Burrows-Wheeler transform (BWT)~\cite{burrows1994block} of the input data and some auxiliary data structures, \hlred{offering two key advantages: i) high compressibility, as the BWT tends to group together characters which appears in similar text strings, and ii) support for arbitrary-length query strings, via the backward search algorithm of BWT.} 
\hlred{These features align well with the requirements of substring cardinality estimation in DBMSs, where both space efficiency and flexible query capabilities are critical.}

However, substantial extensions to the FM-Index are required to effectively support substring cardinality estimation in DBMS. First, FM-Index is primarily designed for indexing a single long string, like a DNA sequence~\cite{das2022memory,simpson2010efficient,li2014fast}, while DBMS needs to process multiple various length strings. One straightforward approach is to process these strings in parallel, but this significantly increases the time required for cardinality estimation.

Second, the extra structure in the FM-Index actually consumes huge space cost, especially when the underlying alphabet set is large~\cite{ferragina2004alphabet,gog2017csa++}. It divides the BWT into fixed-length segments and introduces additional tables to record the cumulative occurrences of each character up to the current offset, eliminating expensive top-down scans. We can see that all characters, whether frequent or rare, need to be cached in each segment. Although such an issue is not serious in bioinformatics as there are only 4 characters, these costs obviously cannot be ignored when the alphabet set is large. We notice that for the rare occurrence of a character $c$, the cached data of $c$ in different tables are similar, leading us to convert the flat-organized segments into tree-organized segments. Thus, cumulative occurrences for rare characters can be recorded in the segment linked from the high-level tree node, which may greatly reduce the space for the extra structure.

Third, the FM-Index can be further reduced in the context of the cardinality estimation. A key point is that the estimated cardinality could accept approximate results, which inspires us to use approximation methods to fit the linear structure in the BWT. For example, we can leverage learned functions with error bounds \cite{galakatos2019fiting,kipf2020radixspline, ferragina2020pgm} to further compress the BWT.

In this paper, we attempt to combine the advantages of the suffix tree and the learning method to extend FM-Index over string columns in the database. Our primary contributions are as follows:

\begin{itemize}

\item We propose SSCard, a \textbf{S}ub\textbf{S}tring \textbf{Card}inality estimator, including i) an enhanced comparator to extend FM-Index to multiple data strings setting naturally; iii) a pruned suffix tree to organize the segments in a hierarchical way as well as a pushup operation to handle the rare and frequent characters adaptively; ii) a learned representation with an error bound to reduce the space cost.

\item We devise several operations on SSCard including: i) a space and time efficient method to construct SSCard; ii) a bidirectional cardinality estimation method that combines the advantages of suffix tree and FM-index; iii) incremental update strategies to handle data changes. We perform the time, space, and error bounds of cardinality estimation.

\item We conduct extensive experiments based on five datasets and report results of accuracy, construction time, estimation time, and space cost compared to existing state-of-art approaches.

\end{itemize}

The rest of the paper is organized as follows. Section~\ref{sec:preliminary} reviews the basic idea of the FM-Index, 
and then formulates our problem. Then we extend the FM-Index to multiple strings, and describe the basic structure of SSCard in Section~\ref{sec:SSCard}, followed by its construction method in Section~\ref{sec:construct}, and cardinality estimation as well as incremental update in Section~\ref{sec:cardinality}. Section~\ref{sec:exp} reports experimental results. The related work is reviewed in Section~\ref{sec:related}, and finally we make a conclusion and foresee the future work in Section~\ref{sec:conclude}.

\section{Preliminary}
\label{sec:preliminary}

We first review the preliminary knowledge of the FM-Index and BWT, which is the basis for the following discussion, and then formulate the problem. The symbols frequently used in this paper are shown in Table \ref{tab:symbols}.

\subsection{FM-Index and BWT}
\label{subsec:bwt}

\begin{figure}[h]
    \includegraphics[width=0.8\columnwidth]{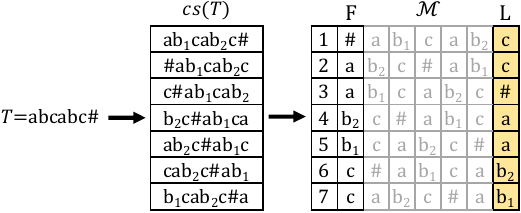}
    \caption{Burrows-Wheeler Transform.}
    \label{fig:bwt}
\end{figure}

FM-Index~\cite{ferragina2000opportunistic} is a string index which can count or locate the occurrences of a substring pattern in a text, and BWT is the basic concept of the FM-Index. As shown in Figure~\ref{fig:bwt}, after placing a special symbol $\#$  ($\#$ is smaller than any character in $\Sigma$) at the end of the text string $T=$ \texttt{abcabc}, BWT creates all its cyclic shifts $cs(T)$ \hlblue{(rotating the first character to the end $|T|$ times)} and sorts them in lexicographical order, forming a conceptual matrix $\mathcal{M}$~\cite{burrows1994block}.

The output of BWT are the $F$ and $L$-array, which are extracted from the first and last columns of $\mathcal{M}$. Operations on the BWT do not rely on the middle parts (gray area in Figure~\ref{fig:bwt}), which can be discarded directly. The $F$ and $L$-array have the following two properties:

\begin{enumerate}

    \item In the $i$-th row, $L[i]$ represents the preceding character of $F[i]$ in $T$, e.g. $T=\cdots L[i]F[i]\cdots$.

    \item The $i$-th occurrence of a character $c$ in the $L$-array is the same as the $i$-th occurrence of $c$ in the $F$-array.

\end{enumerate}

Property 1 is straightforward, while Property 2 is detailed in Figure~\ref{fig:bwt}. To clarify the ordering of the character \texttt{b} in the text string $T$, we add subscripts to distinguish its two occurrences, labeled as \texttt{b}$_1$ and \texttt{b}$_2$. In the $F$-array, \texttt{b}$_2$ appears before \texttt{b}$_1$, which is consistent with their order in the $L$-array. This consistency arises from the same sorting context: the substring following a character in the $F$-array is the same as the substring preceding the character in the $L$-array. More formal proof can be found in \cite{burrows1994block}. This property is the key to counting the occurrences of patterns in $T$.

The BWT-based pattern counting is based on two core functions, in which $Rank(c, i)$ returns the count of occurrences of $c$ in $L[1\dots i]$, and $Occ(c)$ is the position of the first occurrence of $c$ in $F$. \hlblue{Take Figure}~\ref{fig:bwt}\hlblue{as an example}, $Rank($‘$b$’$,7)=2$, meaning that \texttt{b} occurs twice in $L[1\dots 7]$, and $Occ($‘$c$’$)=6$, since \texttt{c} first appears in $F[6]$.

The occurrences of a pattern $P$ can be counted with these two structures in a backward manner~\cite{ferragina2000opportunistic}. 
Take 
an example of pattern $P=$ \texttt{abc}. We first locate the index range $[start_c, end_c]$ for \texttt{c} in the $F$-array. As the same characters are consecutive in the $F$-array, $start_c=6$ and $end_c=7$ can be computed efficiently. According to property 1, the character preceding \texttt{c} is stored in the same row of the $L$-array, which means that the occurrences of \texttt{bc} in $T$ are the same as the occurrences of \texttt{b} in $L[6\dots7]$. 
This can be calculated by $L.Rank($‘$b$’$,6)=1$, and $L.Rank($‘$b$’$,7)=2$.
Then, to count the occurrences of \texttt{abc}, we can apply property 2 to locate the index range of \texttt{bc} in the $F$-array. That is, $start_{bc}=4$ and $end_{bc}=5$ as $start_{bc}=Occ($‘$b$’$)+L.Rank($‘$b$’$,6)-1$ and $end_{bc}=Occ($‘$b$’$)+L.Rank($‘$b$’$,7)-1$.  
Hence we further detect \texttt{a} in $L[4\dots5]$, and two occurrences can be found by taking the same process above. 
Finally, we can conclude that there are 2 occurrences of the pattern $P=$ \texttt{abc} in $T$. The time complexity for counting the occurrences of the pattern $P$ is $O(|P|)$, assuming constant-time implementations of both $Rank(c, i)$ and $Occ(c)$.

The FM-index implements $Rank(c, i)$ and $Occ(c)$ efficiently, but with relatively high space cost. 
$Occ(c)$ can be compressed into a $|\Sigma|$-sized array as the characters in the $F$-array are lexicographically sorted. For $Rank(c,i)$, a straightforward approach requires scanning all preceding characters. In order to lower the scanning cost, the FM-Index divides $L$ into large segments of size $\log^2|T|$, each of which is further divided into smaller segments of size $\log|T|$. With the cumulative occurrences recorded in additional tables for these two segments, $Rank(c,i)$ can be implemented in $O(1)$ time. 

\subsection{Problem Formulation}

Given an underlying data string set $\mathcal{D}$ over an alphabet $\Sigma$, and any pattern string $P$ $(P\in\Sigma^*)$, the cardinality of $P$ returns the number of strings in $\mathcal{D}$ that contain $P$ as a substring. Note that, unlike counting total occurrences in a single string, each string $D \in \mathcal{D}$ is counted at most once in the context of a DBMS, even if $P$ appears multiple times in $D$.

\section{SSCard}
\label{sec:SSCard}

In this section, we first discuss how to extend the FM-Index to multiple underlying data strings, and then present the data structure of SSCard.

\subsection{Extension of the FM-Index to Multiple Strings}
\label{subsec:extend}

The extensions to multiple strings are not straightforward. One simple method is to apply BWT to all strings concatenated with special symbols. However, such a method requires high construction cost in sorting suffixes, and the final $F$ and $L$-array contains unnecessary patterns crossing special symbols. Another option is to leverage BWT for each string separately. The major issue lies in the fact that each FM-Index created by a single string has to be scanned during estimation, leading to high cost even with parallelization. 

\begin{figure}[h]
    \centering
    \includegraphics[width=0.7\linewidth]{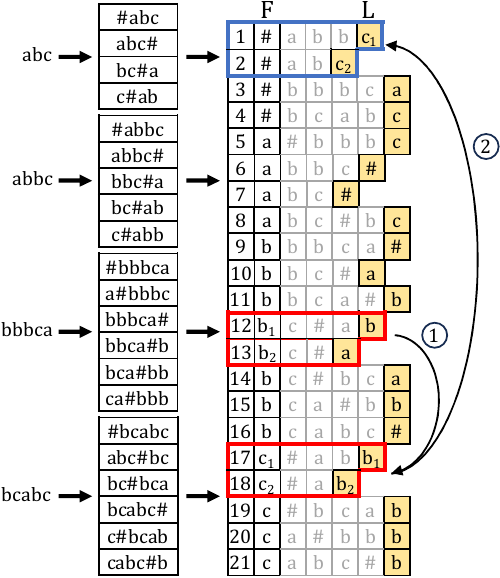}
    \caption{Extension of the BWT to multiple strings.}
    \vspace{-0.5cm}
    \label{fig:multi_bwt}
\end{figure}

Here, we expect to extend BWT on multiple strings naturally. That is, we first build all the cyclic shifts $cs(D_i)$ for each string $D_i \in \mathcal{D}(1\leq i\leq |\mathcal{D}|)$ separately, and then merge them together, resulting in a sorted list. As shown in Figure~\ref{fig:multi_bwt}, the $F$-array and $L$-array are derived from the first and last characters of the list, respectively. Unlike in the case of the single string, \hlblue{the sorted cyclic shifts may have varying lengths,} which requires us to examine whether the two properties discussed in Section \ref{subsec:bwt} still hold. 

\begin{table}[t]
\caption{Frequently used symbols}
\vspace{-0.3cm}
\begin{tabularx}{\linewidth}{|c|X|}
\hline
Symbols &  Meaning \\
\hline

$\mathcal{D}$ & a data string set $\mathcal{D}$ \\

$\mathcal{P}$ & a pattern string set $\mathcal{P}$ \\

$n$ & the number of strings in $\mathcal{D}$, same as $|\mathcal{D}|$\\

$m$ & the maximum length of strings in $\mathcal{D}$\\

$P$, $D$ & a pattern $P\in \mathcal{P}$ against a data string $D\in \mathcal{D}$ \\

$\Sigma$ & the alphabet set $\Sigma$ \\

$h$ & the height of the pruned suffix tree \\

$l$ & the length of a segment linked to a suffix tree node \\

$c_m$ &the minimum \# of identical-character $L$-triples in a segment\\

$\epsilon$ & error bound \\

\hline
\end{tabularx}
\label{tab:symbols}
\end{table}

It is easy to see that Property 1 still holds, as in the $i$-th row, $F[i]$ and $L[i]$ are the first and last characters of the same rotation, but Property 2 may not always be preserved. We can see an example in Figure~\ref{fig:multi_bwt}. \hlblue{In the $12$-th and $13$-th rows, there is no explicit lexicographical order between the string \texttt{bc\#a} and \texttt{bc\#ab}. When we place these strings randomly, they obviously violate Property 2. A similar issue arises with \texttt{c\#abb} in the $17$-th row and \texttt{c\#ab} in the $18$-th row. These cases indicate that lexicographical sorting alone is insufficient to preserve Property 2.

Before delving into the details, we introduce a sorting order between two strings. Without loss of generality, let $T_1$ and $T_2$ be two strings with $T_1$ shorter than $T_2$. $T_1$ and $T_2$ follow the \textbf{lexicographical order} if $T_1$ is not the prefix of $T_2$. Otherwise, if $T_1$ is a prefix of $T_2$, we iteratively rotate both strings simultaneously $k$ times to obtain $T_1'$ and $T_2'$ until $T_1'$ is no longer a prefix of $T_2'$ for the first time. 
The lexicographical order between $T_1'$ and $T_2'$ is then defined as the \textbf{first rotation-based lexicographical order} of $T_1$ and $T_2$. It is easy to know that $k$ is limited, since we place a special symbol $\#$  ($\#$ is smaller than any character in $\Sigma$) at the end of the each string, and $T_1$ and $T_2$ have different lengths.

To satisfy Property 2, the substrings following a character $c$ in the $F$-array must preserve the same order as the substrings preceding $c$ in the $L$-array. We therefore focus on the case where two strings begin with the same character $c$. Let $T_1 = cA$ and $T_2 = cB$, where $A$ and $B$ are the substrings following $c$ in $T_1$ and $T_2$, respectively. 
The problem then reduces to finding an enhanced comparator, in which the order of $cA$ and $cB$ matches that of $Ac$ and $Bc$. We can see that when this condition holds, the order of the occurrences of $c$ in $F$-array is consistent with that in the $L$-array.

We perform a detailed analysis between two strings $T_1=cA$ and $T_2=cB$, in which the relationship between $A$ and $B$ must fall into one of the following cases:
}

\begin{enumerate}

    \item \hlblue{$A$ is not a prefix of $B$: In this case, the lexicographical order between $cA$ and $cB$ is consistent with that between $Ac$ and $Bc$. For example, the $17$-th and $19$-th rows can be sorted explicitly, and their order is consistent with that between the $1$-st and $4$-th rows.
    }

    \item \hlblue{$A$ is a prefix of $B$, but not equal to $B$: In such a case, we apply the first rotation-based lexicographical order of $A$ and $B$. For example, the order of the $12$-th and $13$-th rows can be determined by that of the $17$-th and $18$-th rows, which can be determined further by the $1$-st and $2$-nd rows. The rationale can be explained in reverse: the $1$-st and $2$-nd rows are directly comparable lexicographically. To preserve Property 2, the starting characters \texttt{c}$_1$ and \texttt{c}$_2$ in the $17$-th and $18$-th rows should follow the same order as the ending characters in the $1$-st and $2$-nd rows. Likewise, the $12$-th and $13$-th rows are ordered in the same manner.}

    \item \hlblue{$A$ is the same as $B$: This case occurs when the two data strings are exactly the same. As the lexicographical order can place these strings contiguously in both the $F$ and $L$-array, such a case does not violate Property 2.}

\end{enumerate}

\hlblue{From the above analysis, we conclude that the enhanced comparator---combining the first rotation-based lexicographical order for prefix cases with the standard lexicographical order for other cases---allows us to correctly handle cyclic shifts of all data strings, ensuring compliance with Property 2. This approach offers a practical way to convert variable length strings into an extended BWT, on which both properties hold and standard FM-Index operations can be performed. We leave the detailed construction of the extended BWT in the following section.
}

\begin{figure*}[t]
    \centering
    \includegraphics[width=\textwidth]{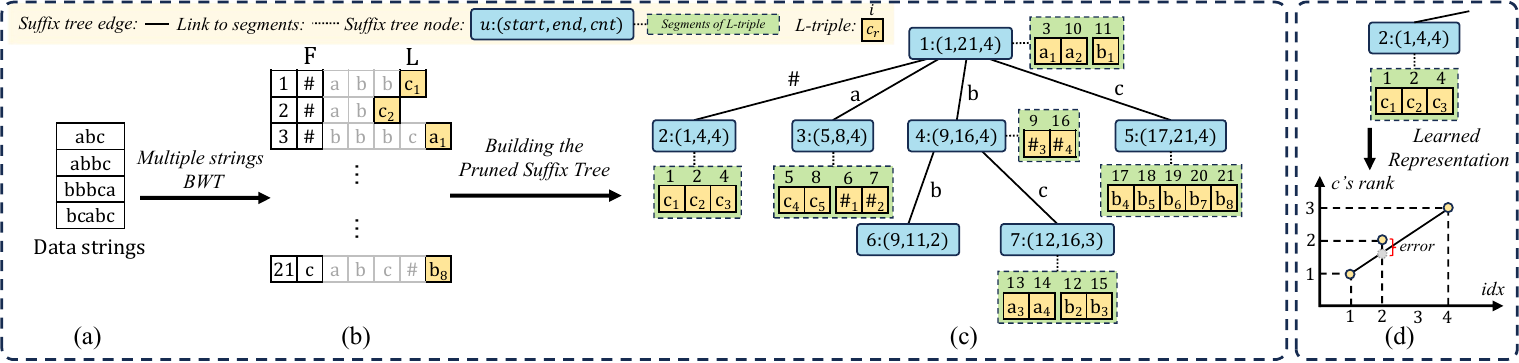}
    \vspace{-0.7cm}
    \caption{\hlblue{An illustration of SSCard with $h$=2 and $c_m=2$ (The meanings of the symbols are in Table}~\ref{tab:symbols}\hlblue{)}}
    \label{fig:architecture}
\end{figure*}

\subsection{Overview of SSCard}
\label{subsec:overview}

SSCard proposed in this paper is based on the FM-Index but employs different auxiliary data structures. We illustrate SSCard in Figure~\ref{fig:architecture} using the same example in Figure~\ref{fig:multi_bwt}. We can see that a pruned suffix tree (a tree with a height restriction) is used to represent $F$ and also organizing the segments of $L$ (Figure~\ref{fig:architecture}c). The segments are further represented by learned functions (Figure~\ref{fig:architecture}d).

Before delving into the structure for the $F$ and $L$-array, we first define the $i$-th $L$-triple. Recall that the $L$-array in BWT is an $O(nm)$-sized character array, which implicitly reveals $(c, i, r)$ indicating the character $c$ in the $i$-th row being the $r$-th occurrences. The introduction of the $L$-triple carries the information explicitly, thereby eliminating the need to store characters in the $L$-array sequentially and allowing them to be put in different places in the following.

\begin{defn}
\label{def:lnode}
\textbf{the $i$-th $L$-triple.} Given the $L$-array in BWT. The $i$-th $L$-triple takes the form of $(L[i],i, r)$, where $r$ is the rank (i.e., the number of prior occurrences) of $L[i]$ in the $L$-array.
\end{defn}

We next describe the structure of the pruned suffix tree for the $F$-array, the pushup operation for $L$-triples in the pruned suffix tree, and the learned representation for the $L$-array.

\stitle{Pruned Suffix Tree for $F$.} In the FM-Index, the middle characters between the $F$-array and $L$-array are removed, and the $F$-array can be further compressed into a $|\Sigma|$-sized array. Through space efficiency, the $F$-array contains limited information. Since rows sharing the same prefix are adjacent in $\mathcal{M}$, we can use a suffix tree to represent $\mathcal{M}$. However, a complete suffix tree takes up too much space. Thus, we construct a pruned version taking only the first few columns of $\mathcal{M}$, subject to three constraints (detailed in Section~\ref{subsec:build_sscard}). In this way, $L$ is divided into segments of $L$-triples and linked to leaf nodes in the suffix tree. 

\hlblue{
The pruned suffix tree not only organizes segments of $L$-triples in a hierarchical way, but introduces another advantage: estimating cardinalities independently and precisely. Specifically, each tree node $u$ can be annotated with the number of strings in $\mathcal{D}$ that match the path from the root to $u$, so as to provide the cardinality for these patterns directly. In addition, we organize the segments by linking them to suffix tree nodes, enabling hierarchical searches and on-demand loading, which reduces access latency to segments.
}

In all, for each node $u$ in the suffix tree, it is annotated with ($start$, $end$, $cnt$, $link$). 
$start$ and $end$ are for the index intervals in the $F$-array (and also the $L$-array) with the same prefix from root to $u$. $cnt$ is for the cardinality of the pattern string from root to $u$, which can estimate the cardinality directly. $link$ is for the link from the suffix tree nodes to the $L$-triple segments. Take Figure~\ref{fig:architecture}c as an example, \hlyellow{$(2,16,3)$} in node $7$ indicates that the index interval in $\mathcal{M}$ \hlyellow{spans from 12 to 16 (inclusive)}, and the cardinality of string $bc$ is $3$. The linked $L$-triples are illustrated in green boxes.

\stitle{\hlblue{The Pushup Operation.}} \hlblue{To improve space efficiency in organizing $L$-triples in the suffix tree, 
we introduce the pushup operation. Since the cumulative occurrences of rare characters may be duplicated across multiple FM-Index segments, the suffix tree provides a chance to construct segments in a hierarchical way to reduce this redundancy. As the $L$-triples explicitly carry all required attributes, rare $L$-triples can be pushed up to the segments linked to higher-level tree nodes, where learned functions are built over consolidated occurrences rather than separately for each one. 
For example, the $16$-th $L$-triple is not linked to the node $7$, but to the node $4$ in Figure}~\ref{fig:architecture}\hlblue{. We will further discuss the pushup operation in Section}~\ref{subsec:build_sscard}.

\stitle{Learned Representation for $L$.}  In contrast to FM-Index, SSCard 
uses functions to represent the distribution of $L$-triples (Figure~\ref{fig:architecture}d). Suppose $S$ be a segment of $L$-triples, SSCard groups $S$ according to different characters in the nodes. For each character $c$, SSCard learns a function $rank_f(c, i)$ to return $r$ as the number of occurrences of $c$ up to $i$. We expect that such a representation can further compress the space requirement for the $L$-array, as multiple $L$-triples are compressed into a single learned function.

Like the FM-Index, SSCard should support $Rank(c,i)$ and $Occ(c)$ efficiently. $Occ(c)$ is actually the $start$ value of the child node reached by following the edge labelled $c$ from the root, 
denoted as $T[c].start$. To compute $Rank(c,i)$, we start from the root and recursively locate the child node $u$ such that $i \in [u.\textit{start}, u.\textit{end}]$, continuing this process until a leaf node is reached. Then run the function $rank_f(c, i)$ stored in the leaf-linked segment $S$.

\section{Construction of SSCard}
\label{sec:construct}

The construction of SSCard is roughly divided into two parts. We first build the pruned suffix tree of SSCard, including locating $L$-triples into the segments linked to tree nodes and the pushup operation. 
We then construct the learned representations for the rank order of $L$-triples. Finally, we conduct the space-time analysis of the construction of SSCard.

\subsection{Building the Pruned Suffix Tree for $F$}
\label{subsec:build_sscard}

We provide a construction method for the data string set $\mathcal{D}$ in Algorithm~\ref{Algorithm:build}. \hlblue{The main function is $Build\_Suffix\_Tree$.} The input parameters are defined as follows. 
The first three parameters are the constraints of the pruned suffix tree, where $h$ limits the maximum height of the tree. The segment size $l$ and the minimum number of $L$-triples containing the same character within a segment $c_m$ prevent constructing small segments with diverse characters. In addition, an error bound $\varepsilon$ is given for the learned representation to balance the estimation accuracy and the space overhead.

First, we need to build the sorted cyclic shifts for all data strings (Figure~\ref{fig:architecture}b), as shown in line 2. We have different choices to achieve the goal. The existing Ukkonen algorithm ~\cite{ukkonen1995line} can construct a suffix tree in $O(m)$ space and time cost for a string of length $m$. Thus, if we concatenate all $n$ strings into a single large one, Ukkonen Algorithm can compute a generalized suffix tree for $\mathcal{D}$ with $O(nm)$ cost. However, the suffix tree generated in this way may contain useless patterns. Even worse, Ukkonen algorithm does not consider cyclic shifts, indicating that the $L$-array has to be located with additional costly searching. In this paper, we adopt a straightforward approach to sort all the cyclic shifts. We apply radix sort on variable-length strings to obtain the sorted cyclic shifts $\mathcal{S}$, using the previously defined comparator. Note that as explicit storage of all cyclic shifts takes $O(nm^2)$ space cost in the worst case, we record the start index and the length to implicitly represent the cyclic shifts in the space cost $O(nm)$.

\begin{figure}[t]
    \centering
    \includegraphics[width=0.9\columnwidth]{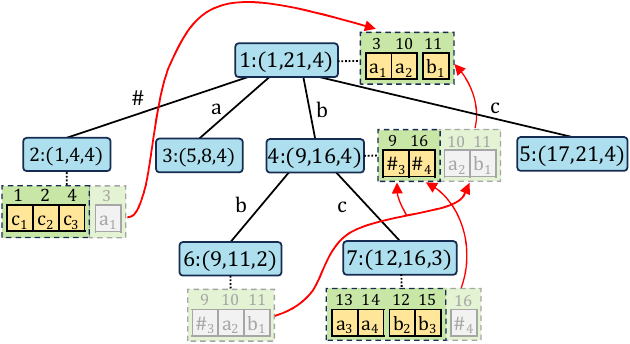}
    \vspace{-0.2cm}
    \caption{\hlblue{An illustration of pushup with $h=2,l=5$ and $c_m=2$. The $L$-triples in node 6 are pushed up to node 4 as their occurrences are smaller than $c_m$. Similar cases apply to the other $L$-triples shown in gray.}}
    \label{fig:pushup}
\end{figure}

In line 3, we locate the sequence $[x[1],\dots,x[nm]]$, where each $x[k]$ is an $L$-triple of the form $(c, i, r)$. This sequence can be located by one-time scan with $O(nm)$ time cost, as the rank order $r$ for each $L$-triple can be computed using an extra $|\Sigma|$-sized structure in the scan. \hlblue{As shown in Figure}~\ref{fig:architecture}\hlblue{b, the $L$-triples are $[(\texttt{c}, 1, 1), (\texttt{c}, 2, 2), \dots, (\texttt{b}, 21, 8)]$.}

\hlblue{
Line 4-7 describe the procedure of constructing the pruned suffix tree $T$. In line 4, we scan $\mathcal{S}$ for $h$ times, in which the $i$-th level of the suffix tree is constructed in the $i$-th round. 
At each level, consecutive chars in $\mathcal{S}[i]$ form the nodes in this layer. These character groups also partition the $L$-array into segments, which are linked to the corresponding suffix tree nodes and further subdivided in the next round. 
To control the space cost, we keep a suffix tree node only if its linked segment contains at least $l$ $L$-triples with each identical character occurring at least $c_m$ times. Lines 5-7 then perform a one-time pre-order traversal to record the start/end indexes of the $L$-array for each tree node and to initialize an $|\Sigma|$-sized empty bucket list $u.B$ for building learned functions. We also annotate the frequency of each prefix at this stage for direct cardinality estimation. Take Figure}~\ref{fig:architecture}\hlblue{c as an example, node 1 is divided into $4$ children, corresponding to $4$ groups of consecutive chars of $\mathcal{S}[1]$. However, only node 4 retains its children as its linked segment contains at least $5$ $L$-triples.}

Lines 9 to 18 outline the framework of pushup operation. We recursively travel the tree and push the infrequent $L$-triples upward to the segments of the father nodes. In lines 10-15, if a node is a leaf, we group $L$-triples from $u.start$ to $u.end$ into buckets according to the character class. And the characters in $L$-triples will be labeled to indicate that they need to be represented by the learning functions (line 14). After that, the pushup operation will decide whether to build functions on these buckets or push them to father nodes (line 15). Because this procedure is recursively executed, we will also perform the pushup operation in non-leaf nodes (line 18).

\begin{algorithm}[t]

\caption{Building SSCard}
\label{Algorithm:build}

\LinesNumbered

\KwIn{String Set $\mathcal{D}$, the maximum height $h$ and segment size $l$ of the pruned suffix tree, the minimum number $c_m$ of $L$-triples with the same class in a segment, error bound $\epsilon$}

\KwOut{SSCard}

\SetKwProg{Fn}{Function}{:}{end}

\Fn{Build\_Suffix\_Tree($\mathcal{D}$, $h$, $l$)}{

Build all cyclic shifts $\mathcal{S}$ from $\mathcal{D}$ using \hlblue{the enhanced comparator mentioned in Section}~\ref{subsec:extend}

Derive the $L$-array using last characters in $\mathcal{S}$

Construct an ordered suffix tree $T$ from $\mathcal{S}$ with the maximum height $h$, and build suffix tree node $u$ when $u$'s segment contains at least $l$ $L$-triples

\For{$u \in T$ in a pre-order visit}{
Record $u's$ start/end indexes in the $L$-array, and the cardinality $u.cnt$

Initialize an $|\Sigma|$-sized 
empty bucket list $u.B$ for $u$
}

Traverse\_Suffix\_Tree($T.root$, $c_m$, $\epsilon$)

\Return{$T$}

}

\SetKwProg{Fn}{Function}{:}{end}

\Fn{Traverse\_Suffix\_Tree($u$, $c_m$, $\epsilon$)}{
\If{$u$ is leaf}{
    \For{$i$ \textnormal{from} u.start \textnormal{to} u.end}{
        Put the $i$-th $L$-triple to bucket $u.B[L[i]]$
    }
    \For {$c \in \Sigma \land u.B[c] \neq \emptyset$}{
        $u.needBuild \leftarrow u.needBuild \cup \{c\}$
    }
    Algorithm~\ref{Algorithm:pushup}: Pushup($u$, $c_m$, $\epsilon$)
}
\For{$v \in u.children$}{
    Traverse\_Suffix\_Tree($v$, $c_m$, $\epsilon$)
}
Algorithm~\ref{Algorithm:pushup}: Pushup($u$, $c_m$, $\epsilon$)

}
\end{algorithm}

Algorithm~\ref{Algorithm:pushup} provides the details of pushup. For each character $c$, we further check whether there are at least $c_m$ $L$-triples in $u.B[c]$ (lines 2-3) if the $L$-triples of $c$ need to be fitted. If both conditions are satisfied, we
build functions to represent $u.B[c]$ (lines 4-5). 
Otherwise, the $L$-triples are pushed up to the bucket of the father node $u.fa.B[c]$. Note that the distribution of $u.B[c]$ also needs to be pushed up in order to build functions for the sibling nodes. Since the functions of these nodes have already been constructed, only the first and last $L$-triples in $u.B[c]$ need to be pushed up, i.e., a compression form (line 5,9). Take Figure~\ref{fig:pushup} as an example, the $9$-th to $11$-th $L$-triples in node $6$ are pushed up to node $4$ because they all appear once. Since the occurrences of the $10$-th, $11$-th $L$-triples in node $4$ are still smaller than $c_m$, they need to be further pushed up to the root node. After node $u$ is pushed up, the bucket and the labels of $u$ are no longer needed (line 10).

\subsection{Building the Learned Representation for $L$}
\label{subsec:fit_L}

This subsection discusses the details of building fitting functions for buckets. Let $x=(c, i, r)$ be \hlyellow{an $L$-triple}, $B=[x[1], \dots, x[k]]$ be the bucket of a $L$-triple sequence with the same character class $c$ in a segment. The learning representation for $L$ is to find a function $rank_f(c, x[j].i)$, whose result is close to $x[j].r$. The problem is similar to that in the learned index~\cite{galakatos2019fiting,kipf2020radixspline}, where $r$ can be seen as the cumulative distribution function over $i$. We use the following two strategies.

\begin{algorithm}[t]
\caption{Pushup}
\label{Algorithm:pushup}
\LinesNumbered

\KwIn{Suffix tree node $u$, the minimum number $c_m$ of $L$-triples with the same class in a segment, error bound $\epsilon$}

\For{$c \in \Sigma$}{

    \eIf{$c \in u.needBuild$}{
        \eIf{$|u.B[c]| >= c_m$}{
            $u.rank_f(c) \leftarrow$ build fitting functions of $u.B[c]$ with $\epsilon$
            
            push the first and last $L$-triples in $u.B[c]$ to $u.fa.B[c]$
        }{
            push all $L$-triples in $u.B[c]$ to $u.fa.B[c]$
            $u.fa.needBuild \leftarrow 
            u.fa.needBuild \cup \{C\}$
        }
    }{
        push the first and last $L$-triples in $u.B[c]$ to $u.fa.B[c]$
    }
}
\textbf{delete} $u.needBuild, u.B$



\end{algorithm}

\stitle{Linear Fitting for $rank_f(c, x[j].i)$.} SSCard can perform linear fitting over the bucket $B=[x[1],\dots,x[k]]$ 
with the objective of minimizing the mean squared error (MSE). For each $L$-triple in a bucket, the linear function takes the following form, where $k_c$ is the slope and $b_c$ is the intercept:
\begin{equation}
rank_f(c, x[j].i)=k_c \times x[j].i + b_c
\end{equation}
The loss functions are as follows:
\begin{equation}
\label{eq:mse}
L = \sum_{j \in |B|}{(rank_f(c, x[j].i)- x[j].r)^2}
\end{equation}
The parameters can be efficiently optimized using the least squares method. \hlred{When the learning convergences}, we can discard the explicit $L$-triple directly and keep only two parameters $k_c$ and $b_c$ for each bucket (character class) in a segment.

\stitle{Greedy Spline Interpolation for $rank_f(c, x[j].i)$.} Another approach to represent these $L$-triples takes the linear spline interpolation~\cite{ferragina2020pgm, galakatos2019fiting, kipf2020radixspline, neumann2008smooth}. There are two differences from the previous one. First, MSE as an objective function can lead to poor fitting results when there are minor changes in $B=[x[1],\dots,x[k]]$~\cite{neumann2008smooth}. Second, the previous method cannot ensure that the start node and the end node match exactly the results of the learned function, which may lead to large errors in the cardinality estimation. Thus we introduce another loss function~\cite{neumann2008smooth}
\begin{equation}
\label{eq:mae}
L = \max_{j \in |B|}\{rank_f(c, x[j].i)- x[j].r\}
\end{equation}
as the Absolute Maximum Error (MAE), and this function may produce more stable training.

We take the spline interpolation mentioned in ~\cite{neumann2008smooth} over $B=[x[1], \dots, x[k]]$ to address these issues. The spline starts from the first $L$-triple $x[1]$, and the linear function $k_c \times (x[j].i-x[1].i) + x[1].i$ can produce the exact matching at $x[1]$. Then the lower and upper bounds of the slope $k_c$ are computed to handle the next $L$-triples with the given error bound. The slope bound of $k_c$ is further adjusted if the following $L$-triples can be represented by the same spline satisfying the error bound. Otherwise, the current spline is terminated, and another new spline is initialized. 
We can see that multiple linear splines may be required to represent $L$-triples for the same character class in one segment, and the given error bound balances space consumption and the quality of cardinality estimation. Such a greedy method can generate the interpolation using a one-time scan of the $L$-triple sequence $B$.

\subsection{Analysis}

We perform the analysis of the SSCard construction method in terms of time/space complexity, using the symbols in Table~\ref{tab:symbols}.

\stitle{Time Complexity in Construction.} The time complexity for the major steps in building the SSCard is shown as follows: We have $O(nm)$ cyclic shifts with the implicit index representation for each cyclic shift. The radix sort takes $O(nm^2)$ in the worst case. The location of the $L$-triple sequence costs $O(nm)$ time. The construction of the pruned suffix tree takes at most $O(|\Sigma|^h)$ in the worst case. The pre-order traversal of the pruned suffix tree is equal to a one-time scan of the $L$-array using $O(nm)$ time cost. 
Applying pushup recursively and constructing the learned representation are also equivalent to a one-time scan of the $L$-triples. In total, the construction of the SSCard takes $O(\max(nm^2, |\Sigma|^h))$ time cost. Usually, $h$ is not large enough, and the time complexity in the construction of SSCard is reduced to $O(nm^2)$ in the worst case.

\stitle{Space Complexity in Construction.} With the similar steps above, we perform analysis over the space complexity. All cyclic shifts (including the sorted results) take $O(nm)$ space, as each one is represented by the start index and the string ID. The pruned suffix tree takes at most $O(|\Sigma|^h)$ space cost as we put the restriction on the tree height. SSCard needs not to store all $L$-triples explicitly but the learned functions, which takes the space cost $O(nm/c_m)$ in the worst case. In total, except for the intermediate space required in the building process, the space cost of SSCard takes $O(nm/c_m)$, as the pruned suffix tree usually takes less space with a small $h$ (\eg 3).

\section{Operations on SSCard}
\label{sec:cardinality}

This section proposes a bidirectional method over SSCard to estimate the pattern cardinality, discusses dynamic updates on SSCard, and then analyzes the complexity and error bounds of cardinality estimation.

\subsection{Bidirectional Estimation}

SSCard can be viewed as two pre-computed structures: i) The pruned suffix tree, which records the cardinalities of strings with lengths smaller than $h$; ii) the learned FM-Index, which supports approximate cardinality estimation for arbitrary-length query strings. We aim to develop a combination of these two structures to fully leverage the exact cardinality information from the suffix tree, even when the pattern length exceeds the tree height.

It is not easy to combine the results of the suffix tree and the FM-Index together as the pattern matching directions are different. For the suffix tree, we need to perform a forward search, while we perform a backward search using the FM-Index. However, we cannot directly combine the results from two directions, as the character occurrences at the $i$-th layer ($i\geq 2$) do not follow the same order as those in the $F$ and $L$-array, and consequently, the intermediate results cannot further guide the next backward search.

\begin{figure}[h]
    \centering
    \hspace{0.03\columnwidth}
    \begin{subfigure}[b]{0.35\columnwidth}
        \includegraphics[width=\linewidth]{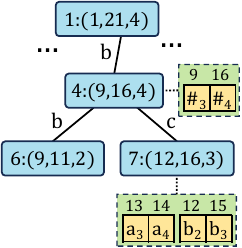}
        \caption{Partial SSCard}
        \label{fig:estimate1}
    \end{subfigure}
    \hfill
    \begin{subfigure}[b]{0.55\columnwidth}
        \includegraphics[width=\linewidth]{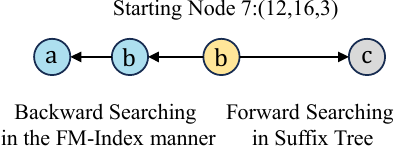}
        \vspace{0.01\textheight}
        \caption{Bidirectional Estimation}
        \label{fig:estimate2}
    \end{subfigure}
    \hspace{0.03\columnwidth}
    \vspace{-0.3cm}
    \caption{\hlblue{Cardinaltity estimation using SSCard}}
    \label{fig:estimate}
\end{figure}

We attempt to combine two structures from another viewpoint. For a pattern $P$, we start with a middle char $P[i](1\leq i \leq |P|)$ rather than $P[0]$, where the suffix started from $P[i]$ can be fully matched in the suffix tree. This also indicates that there is a consecutive index interval for the suffix $P[i]$ in the $F$-array, from which the backward search can be further performed in the FM-Index. We can see the basic idea in Figure~\ref{fig:estimate}. The second \texttt{b} is the starting char, since the suffix \texttt{bc} is the longest subpattern that can be matched in the pruned suffix tree in Figure~\ref{fig:architecture}. We can locate the valid index range $[12,16]$ in the $F$-array for \texttt{b}, from which the prefix \texttt{abb} can be further matched with the backward search using the idea of the FM-Index. The results in the backward search are the results for the overall pattern, which have considered both suffix and prefix subpatterns.


\begin{defn}
\label{def:startnode}
\textbf{Starting character in estimation.} Given a pattern string $P$ and a pruned suffix tree $T$, a char $P[i](1\leq i \leq |P|)$ is the starting character, if the suffix started from $P[i]$ matches a path in $T$, and no suffix started from $P[k](k<i)$ matches any path in $T$.
\end{defn}

Before giving the full method in the cardinality deestimation, we first detail the $Rank(c, i)$ used in SSCard in Algorithm~\ref{Algorithm:rank}. Given the $i$-th index and a character $c$, the major step is to locate the segment that stores its learned function $rank_f(c, i)$, which can be used to answer $Rank(c,i)$ directly. As we have performed the pushup operation on rare characters, not all functions are stored in the segment linked by the leaf node in the suffix tree. However, we know that the function must appear in the last segment that contains the function with $c$, according to the pushup rules in the SSCard construction. Thus we first iteratively traverse $T$ to a leaf node by choosing the child node $v$ that satisfies $i \in [v.start,v.end]$ (lines 1-3), and then traceback to the father node of $u$ if $u$ does not contain $rank_f(c, i)$ (lines 4-5). \hlblue{Take $rank_f(\texttt{\#}, 11)$ as an example. After line 3, we reach node 6, but node 6 does not contain a   learned function for \texttt{\#}. 
Therefore, we backtrack to its parent, node 4, which contains the required function, and compute $rank_f(\texttt{\#}, 11) = 3$.}

\begin{algorithm}

\caption{Rank Computation $rank_f(c, i)$}
\label{Algorithm:rank}

\LinesNumbered

\KwIn{Character $c$, the index $i$, SSCard}

\KwOut{Rank order $r$ of $c$ in the $L$-array}

$u \leftarrow T.root$

\While{$u$ is not leaf}{
    $u \leftarrow v$, where $v \in u.children \land i \in [v.start, v.end]$
}

\While {$u.rank_f(c) = \emptyset \land u \neq root$}{
    $u \leftarrow u.fa$
}

Return $u.rank_f(c, i)$

\end{algorithm}

With the introduction of the starting character and implementation of $Rank(c,i)$, we give an overall bidirectional estimation of the SSCard in Algorithm~\ref{Algorithm:Est}. The main idea is similar to the FM-Index, which iteratively computes the index interval for the preceding chars on the $L$-array. 
First, we locate the starting node using a binary search on $P$ and check whether the suffix is matched in the pruned suffix tree $T$ in line 1, as the height of some leaves in $T$ is smaller than $h$. That is, we first choose a suffix from the middle char $P[k]$ and attempt to match the suffix in $T$. If the suffix is not preserved in $T$, we change the starting char to the right suffix in $P$. Otherwise, we change the starting char to the left suffix until the suffix with the maximum length can be located in $T$. Second, we use the approximate function $rank_f(c, i)$ in Algorithm~\ref{Algorithm:rank} to replace the exact $Rank(c,i)$ in lines 5-6, where $T[c_{k-1}].start$ is the implementation of $Occ(c_{k-1})$ mentioned in Section~\ref{subsec:overview}. \hlblue{As illustrated in Figure}~\ref{fig:estimate2}\hlblue{, the starting character of \texttt{abbc} is $P[3]=\texttt{b}$, and the index interval is $[12, 16]$. Then the backward procedure starts from $[12, 16]$ ($start=12$ and $end=16$ in Line 2) and continues to estimate \texttt{abb} in the same way as the FM-Index.}



\begin{algorithm}

\caption{Cardinality Estimation by SSCard}
\label{Algorithm:Est}

\LinesNumbered

\KwIn{A Pattern $P$, SSCard}

\KwOut{Estimated Cardinality of $P$}

Determine the starting character $P[k]$ using the binary search over the pruned suffix tree $T$

Locate the index interval $[start, end]$ for the suffix sub-pattern from $P[k]$ in $T$

\While {$start \leq end$ {\bf and} $k-1\geq 2$}{

    $c_{k-1} \leftarrow P[k-1]$

    $start \leftarrow T[c_{k-1}].start + rank_f(c_{k-1},start)$

    $end \leftarrow T[c_{k-1}].start + rank_f(c_{k-1},end)$

    $k \leftarrow k-1$
}

Return $end - start + 1$

\end{algorithm}

\subsection{Incremental Update}
\label{subsec:update}

\hlblue{I}\hlred{ncremental updates to SSCard are non-trivial. For example, during insertion, the pruned suffix tree lacks full information about the cyclic shifts, making it hard to directly place new $L$-triples into the correct segments. Moreover, inserting a new string alters the rank of $L$-triples, thereby invalidating the existing learned functions.}

\hlblue{S}\hlred{SCard allocates a fixed-size main memory to cache a pruned suffix tree for newly added data strings. That is, in addition to the original SSCard, we maintain a pruned suffix tree $T_i$ specifically for newly added data, as $T_i$ can be efficiently updated and directly used for cardinality estimation. 
For each newly added string, we identify all its suffixes shorter than a threshold $h$ and insert these suffixes into $T_i$, during which the cardinality $u.cnt$ for $u\in T_i$ is updated in the same way as in SSCard discussed before. The results from SSCard and the suffix tree are then combined to estimate the cardinality of a given pattern.
}

\hlblue{W}\hlred{e adopt two strategies when the size of $T_i$ exceeds the fixed memory budeget. The first is the \textbf{single-SSCard update strategy}, which rebuilds a single SSCard by merging the original data strings and inserted data strings when the memory is full. The second, inspired by update strategies used in LSM-Tree families}~\cite{o1996log, ferragina2020pgm, luo2020lsm}\hlred{, is the \textbf{multiple-SSCard update strategy}. Rather than merging all data into one structure, we create an independent SSCard for the newly inserted data. As a result, multiple SSCards coexist and can later be consolidated during system idle periods. The multiple-SSCard update strategy is expected to greatly reduce update costs, at the expense of slightly increased cardinality estimation time and Q-error compared to the Single-SSCard update strategy.} 

The deletion of the old string from SSCard can be supported in a similar way to handle the insertion, by building a new pruned suffix tree $T_d$ to cache the deleted strings.

\begin{table*}[t]
\centering
\caption{Statistics about datasets for evaluation}
\vspace{-0.3cm}
\label{tab:datasets}
\begin{tabular}{|c|c|c|c|c|c|c|c|c|c|}
    \hline
    \multirow{2}{*}{Datasets} & \multicolumn{5}{c|}{String Set $\mathcal{D}$} & \multicolumn{4}{c|}{Query Set $\mathcal{P}$} \\
    \cline{2-10}
    & $|\Sigma|$ & $|\mathcal{D}|$ & $l_{avg}$ & $l_{max}$ & $l_{sum}$ & $|\mathcal{P}|$ & $l_{avg}$ & $l_{min}$ & $l_{max}$ \\
    \hline
    DBLP-AN & 27 & 450,000 & 14.6 & 44 & 6,572,365 & 400,000 & 6.0 & 1 & 8\\
    \hline
    IMDB-AN & 27 & 550,000 & 14.1 & 39 & 7,774,344 & 500,000 & 6.2 & 1 & 8\\
    \hline
    IMDB-MT & 38 & 500,000 & 14.7 & 79 & 7,366,281 & 500,000 & 6.2 & 1 & 8\\
    \hline
    TPCH-PN & 26 & 200,000 & 32.8 & 50 & 6,550,221 & 1,128 & 3.9 & 1 & 8\\
    \hline
    WIKI & 3735 & 1,031,930 & 98.4 & 300 & 101,561,110 & 1,000,000 & \hlgreen{6.1} & 1 & \hlgreen{8}\\
    \hline
\end{tabular}
\end{table*}

\subsection{Analysis}
\label{subsec:analysis}

We first discuss the time complexity and error bounds in the estimation, and then analyze the time/space cost in the dynamic update.

\stitle{Time Cost in Estimation.} The cardinality estimation using SSCard involves two major steps. The location of the starting node requires $O(\log |P|)$ times trials, each of which requires $O(h)$ time cost. Thus, searching over the pruned suffix tree takes $O(h\cdot \log |P|)$. The search over SSCard takes at most $|P|$ steps in the worst case, and in each step, we locate the target segment via at most $h$ access to the suffix tree. The learning function for {\sl Rank} takes a constant time. Thus, the total time cost of searching SSCard is $O(h\cdot |P|)$. By combining two factors, the total time cost is $O(h\cdot |P|)$. In practice, since $h$ is usually a small constant, the overall time complexity can be approximated as linear, i.e. $O(|P|)$.

\stitle{Estimation Error Bound.} The estimation error of SSCard includes two parts: 1) the multiple occurrences of $P$ in a single data string $D \in \mathcal{D}$, and 2) the cumulative error of approximate $Rank(c,i)$ over the $L$-array. In practice, the first issue is serious when the length of a pattern is limited but can be significantly alleviated with $cnt$ annotated to the suffix tree nodes. We mainly focus on the second issue using greedy spline interpolation with the error bound $\epsilon$. In each step, we estimate an index interval using the learned functions $rank_f(c, i)$ two times. In the worst case, the error introduced in one backward step is $2\epsilon$. With the total $|P|$ steps, the absolute error distance to the correct one is at most $2\epsilon\cdot |P|$ in the worst case.

\stitle{Time and Space Cost in Update.} For a newly inserted string $D$, it takes at most $O(|D|^2\cdot h)$ time to add all suffixes to the incremental suffix tree $T$, where $h$ is the maximum height of $T$. The time complexity of rebuilding the SSCard is theoretically the same as building the SSCard from scratch, which takes $O(nm^2)$ time. For the space cost, the incremental suffix tree takes $O(|\Sigma|^h)$ in the worst case, where $h$ is the maximum height of the cached suffix tree.

\section{Experiments}
\label{sec:exp}

We conduct experiments on five real-life datasets, comparing SSCard with the state-of-the-art (SOTA) estimators, and evaluating the effectiveness of different components in SSCard. Our source code is available at https://github.com/marlcplhra/SSCard.

\subsection{Experimental Setup}

 \stitle{Experiment Environment:} SSCard and other competitors are implemented with Python on a machine with Intel(R) Xeon(R) Silver 4210R CPU @ 2.40GHz, and 256 GB of RAM, Ubuntu 20.04.6 LTS. For learning-based methods, a single NVIDIA GeForce RTX 3090 GPU is used for training.

\stitle{Datasets:} We use the same datasets as the existing works~\cite{lee2007extending, shetiya2020astrid, aytimur2024lplm}, 
sampling author names (DBLP-AN) from DBLP\footnote{https://dblp.org}, actor names (IMDB-AN) from IMDB\footnote{https://imdb.com}, movie titles (IMDB-MT) from IMDB and part names (TPCH-PN) from TPCH\footnote{https://www.tpc.org/tpch/}. These four datasets have a small alphabet, thus we include the WIKI dataset which is used in DREAM~\cite{kwon2022cardinality} to analyze method's scalability for larger character sets. 
It is constructed by randomly sampling sentences from 101,873 Wikipedia articles in the DocRED dataset~\cite{yao2019docred}. 
The query set $\mathcal{P}$ is constructed by sampling substrings of words from the string set $\mathcal{D}$. Notice that TPCH-PN has only 92 unique words, thus there are only 1,128 unique substrings in total. The actual cardinalities of pattern strings in $\mathcal{P}$ is generated by using the same summary data structure in~\cite{shetiya2020astrid}, and we take 50\% of $\mathcal{P}$ for training and 50\% as the test set. Within 50\% of the training data, we take 10\% for validation. For estimators that do not require training, we test their performance directly on test sets.

{
\renewcommand{\arraystretch}{1.3} 
\begin{table*}[t]
\centering
\caption{Q-error of cardinality estimation}
\vspace{-0.3cm}
\label{tab:q-error_card}
\setlength{\tabcolsep}{3pt}

\resizebox{\textwidth}{!}{
\begin{tabular}{|c|*{25}{c|}}
\hline
\multirow{2}{*}{Estimators} & \multicolumn{5}{c|}{DBLP-AN} & \multicolumn{5}{c|}{IMDB-AN} &\multicolumn{5}{c|}{IMDB-MT} & \multicolumn{5}{c|}{TPCH-PN} & \multicolumn{5}{c|}{WIKI}\\
\cline{2-26}
 & Avg. & 50th & 90th & 99th & Max. & Avg. & 50th & 90th & 99th & Max. & Avg. & 50th & 90th & 99th & Max. & Avg. & 50th & 90th & 99th & Max. & Avg. & 50th & 90th & 99th & Max.\\
 \hline
 \hlblue{MO} & 3.72 & \textbf{1.67} & 6.75 & 21.7 & 5968.9 & 2.48 & \textbf{1.00} & \textbf{3.79} & 16.55 & 19600.0 & 5.44 & \textbf{1.00} & \textbf{3.14} & \textbf{14.1} & 57451.9 & \textbf{1.00} & \textbf{1.00} & \textbf{1.00} & \textbf{1.00} & \textbf{1.00} & 86.9 & 1.98 & 16.5 & 730.0 & 460337.4 \\
\hline
LBS & \underline{3.12} & 1.98 & 8.00 & \textbf{17.0} & \underline{537.8} & \underline{2.20} & \textbf{1.00} & 5.00 & \underline{15.0} & 2273.0 & \underline{2.28} & \textbf{1.00} & 5.00 & \underline{16.0} & \underline{214.4} & \underline{1.02} & \textbf{1.00} & \underline{1.02} & \underline{1.37} & 2.24 & \underline{3.21} & \textbf{1.02} & \underline{7.00} & \textbf{20.0} & \underline{2372.4} \\
\hline
Astrid & 3.84 & 2.27 & \underline{5.97} & 30.2 & 1930.0 & 3.26 & 1.72 & 4.43 & 28.8 & 1330.6 & 3.23 & 1.75 & 4.74 & 28.9 & 2251.1 & 2.44 & 2.01 & 3.07 & 12.7 & 29.2 & 10.5 & 3.05 & 11.4 & 86.2 & 488367.7 \\
\hline
DREAM & 3.21 & 1.89 & \textbf{5.47} & 22.9 & 1055.5 & 2.68 & \underline{1.35} & \underline{3.95} & 20.0 & 2359.3 & 2.59 & \underline{1.33} & 4.07 & 21.0 & 805.8 & 1.83 & \underline{1.20} & 1.65 & 21.1 & 71.8 & 6.23 & \underline{1.82} & 7.35 & \underline{61.6} & 37514.1 \\
\hline
LPLM & 5.23 & 2.89 & 10.6 & 35.9 & 1342.4 & 3.44 & 2.16 & 6.09 & 22.1 & \underline{1167.2} & 4.62 & 2.45 & 7.43 & 27.9 & 31473.8 & 1.85 & 1.64 & 2.75 & 4.09 & 4.32 & 93.6 & 12.7 & 102.4 & 823.9 & 1016548.9 \\
\hline
SSCard & \textbf{2.92} & \underline{1.80} & 6.00 & \underline{20.0} & \textbf{64.0} & \textbf{2.05} & \textbf{1.00} & 4.00 & \textbf{14.0} & \textbf{70.0} & \hlgreen{\textbf{2.14}} & \textbf{1.00} & \underline{4.00} & \underline{16.0} & \textbf{71.0} & \textbf{1.00} & \textbf{1.00} & \textbf{1.00} & \textbf{1.00} & \underline{1.02} & \textbf{2.61} & \textbf{1.02} & \textbf{5.00} & \textbf{20.0} & \textbf{73.0} \\
\hline

\end{tabular}
}

\end{table*}
}

{
\renewcommand{\arraystretch}{1.3} 
\begin{table*}[t]
\centering
\caption{Runtime of entire pipeline for cardinality estimators. P., T. Q. correspond to pre-processing, training and query time}
\vspace{-0.3cm}
\label{tab:runtime}
\setlength{\tabcolsep}{3pt}

\resizebox{\textwidth}{!}{
\begin{tabular}{|c|*{5}{cc|}}
\hline
\multirow{2}{*}{Estimators} & \multicolumn{2}{c|}{DBLP-AN} & \multicolumn{2}{c|}{IMDB-AN} &\multicolumn{2}{c|}{IMDB-MT} & \multicolumn{2}{c|}{TPCH-PN} & \multicolumn{2}{c|}{WIKI}\\
\cline{2-11}
& P. + T. & Q. (ms) & P. + T. & Q. (ms) & P. + T. & Q. (ms) & P. + T. & Q. (ms) & P. + T. & Q. (ms) \\
\hline
\hlblue{MO} & 2.97 (min) + 0 & \textbf{0.02} & 4.5 (min) + 0 & \textbf{0.02} & 7.5 (min) + 0 & \textbf{0.02} & 6.0 (min) + 0 & \textbf{0.01} & 3.5 (h) + 0 & 0.13 \\
\hline
LBS & 16.6 (min) + 0 & 0.06 & 20.3 (min) + 0 & 0.06 & 19.5 (min) + 0 & 0.07 & 16.7 (min) + 0 & 0.1 & 7.2 (h) + 0 & 0.05 \\
\hline
Astrid & 27.5 (s) + 2.9 (h) & 1.42 & 53.2 (s) + 4.9 (h) & 1.43 & 47.2 (s) + 5.3 (h) & 1.46 & 0.1 (s) + 22.9 (s) & 1.71 & 2.3 (min) + 10.8 (h) & 2.34 \\
\hline
DREAM & 4.8 (s) + 47.4 (min) & 4.4 & 8.1 (s) + 1.5 (h) & 7.9 & 7.2 (s) + 1.4 (h) & 7.7 & \textbf{0.1 (s) + 18.4 (s)} & 6.3 & 25.6 (s) + 2.6 (h) & 4.6 \\
\hline
LPLM & 2.6 (h) + 14.7 (min) & 0.16 & 4.0 (h) + 13.1 (min) & 0.16 & 3.7 (h) + 13.2 (min) & 0.15 & 13.7 (s) + 71.4 (s) & 0.25 & 22.8 (h) + 0.48 (h) & 6.5 \\
\hline
SSCard & \textbf{2.0 (min) + 32.4 (s)} & 0.06 & \textbf{1.3 (min) + 39.5 (s)} & 0.06 & \textbf{1.4 (min) + 37.0 (s)} & 0.06 & 1.1 (min) + 29.3 (s) & 0.02 & \textbf{2.2 (h) + 8.9 (min)} & \textbf{0.04} \\
\hline

\end{tabular}
}

\end{table*}
}

\begin{figure*}[!htb]
    \centering
    \includegraphics[width=\linewidth]{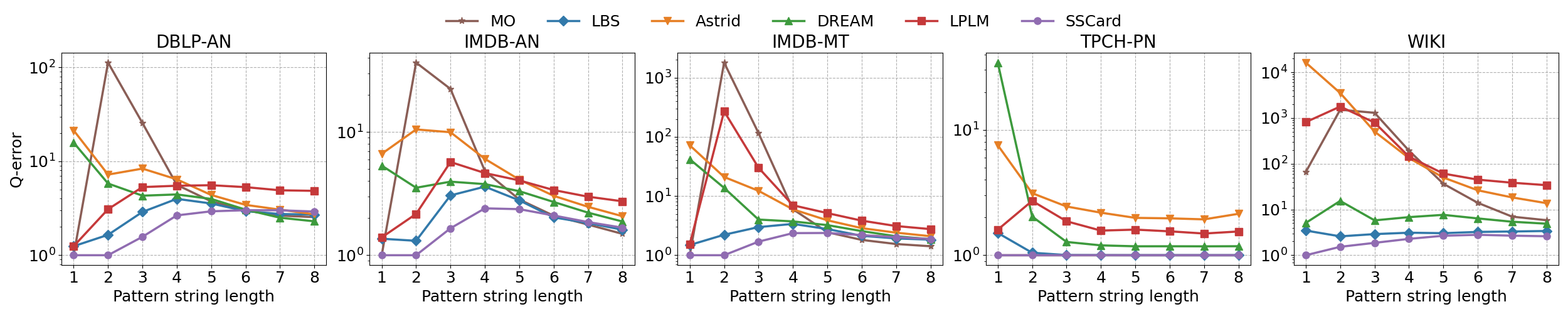}
    \vspace{-0.7cm}
    \caption{\hlblue{Q-error of estimators varying pattern lengths}}
    \label{fig:q_error_l}
\end{figure*}

\begin{figure*}[!htb]
    \centering
    \includegraphics[width=\linewidth]{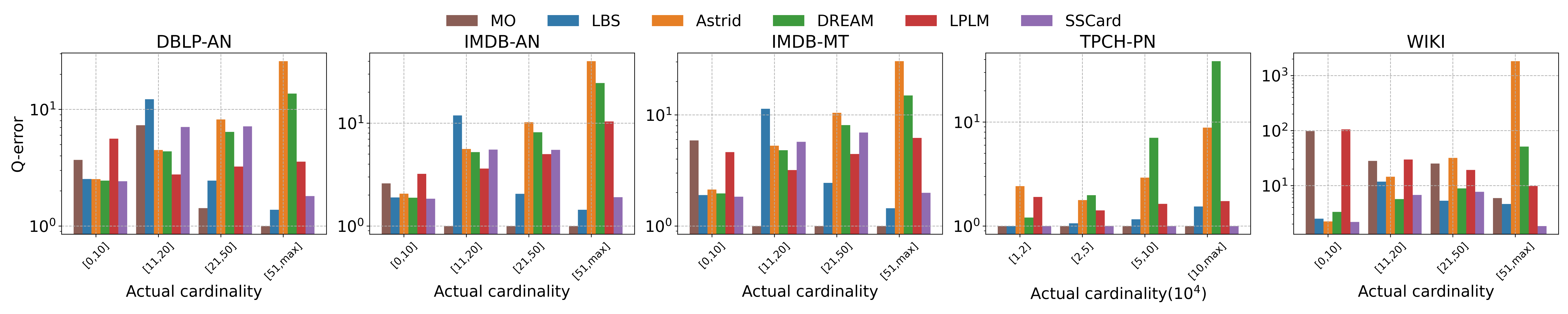}
    \vspace{-0.8cm}
    \caption{\hlblue{Q-error of estimators varying actual result cardinality}}
    \label{fig:q_error_c}
\end{figure*}

\stitle{Evaluation Metrics:} We evaluate the estimation accuracy, construction time, estimation time, and estimator size of SSCard and competitors. For estimation accuracy, we utilize q-error as the evaluation metric, which is widely used in previous work~\cite{shetiya2020astrid,kwon2022cardinality,hilprecht2019deepdb,kipf2019estimating,kipf2018learned}. Q-error is defined as follows, where $y_i$ is an actual value and $\Tilde{y}_i$ is an estimated value,
\begin{gather}
    q-error(y_i,\Tilde{y}_i) = \max(\frac{y_i}{\Tilde{y}_i},\frac{\Tilde{y}_i}{y_i})
\end{gather}
To avoid dividing by zero, the lower bounds of $y_i$ and $\Tilde{y_i}$ are set to $1$ as in other works~\cite{shetiya2020astrid, aytimur2024lplm,kwon2022cardinality}. 
The size of an estimator is measured by the absolute size of the file when the estimator is saved into the secondary memory.

\stitle{Competitors:} We compare SSCard with the following methods:
\begin{itemize}
    \item \hlblue{MO: MO}~\cite{jagadish1999substring} \hlblue{is a substring cardinality estimator based on pruned suffix tree. We set the prune threshold to $2\%$, reserving $2\%$ nodes of the full suffix tree.}
    \item LBS: LBS~\cite{lee2009approximate} is an estimator for approximate string queries based on $k$-gram.
    \item Astrid: Astrid~\cite{shetiya2020astrid} learns selectivity-aware embeddings of substrings from $\mathcal{D}$. 
    We train the embedding model using $\mathcal{D}$ and the selectivity model using the triplets generated from the training set of $\mathcal{P}$.
    \item DREAM: DREAM~\cite{kwon2022cardinality} is the SOTA estimator for approximate string queries. We set the edit distance~\cite{gusfield1997algorithms} threshold to $0$ when comparing with other approaches. We use the base training data of $\mathcal{P}$ to train the model.
    \item LPLM: LPLM~\cite{aytimur2024lplm} is the SOTA estimator for LIKE predicates. We take $\mathcal{P}$ as the training set and utilize 500 SQLite databases \hlyellow{in parallel} to generate the ground truth probabilities of LPLM as mentioned in~\cite{aytimur2024lplm}.
\end{itemize}

\textbf{Hyperparameter:} SSCard has four hyperparameters, including the height of the pruned suffix tree $h$, the minimum number of occurrences for each identical-character in a segment $c_m$, the segment size $l$, and the error bound $\epsilon$ for spline interpolation. 
We set $h=3,c_m=10,l=5000$ and $\epsilon=32$ for all five datasets.


\subsection{General Comparison}


\stitle{Estimator Accuracy:} Table \ref{tab:q-error_card} presents the q-error results of SSCard and other approaches by the average value, the values at the $50$-th, $90$-th, $99$-th percentiles, and the maximum value. 
Bold numbers indicate the best results, while underlined numbers indicate the second-best results. Results show that SSCard outperforms other competitors in terms of q-errors from all aspects. 


We compare SSCard with the second-best methods in terms of q-error. On DBLP-AN, IMDB-AN and IMDB-MT datasets, SSCard achieves an approximate $15\%$ reduction on average, as the moderate number of data strings and a relatively small alphabet size enable accurate learned representation within error bounds. \hlblue{While MO demonstrates low q-error at the $50$-th and $90$-th percentiles, it performs poorly in terms of the maximum q-error. This is primarily due to the character-wise independent estimation of short strings in MO, which often leads to substantial overestimation, as also shown in Figure}~\ref{fig:q_error_l}. LBS achieves similar performance to SSCard in TPCH-PN due to the short pattern strings, with the $N$-gram table (maximum length $5$) covering most queries. DREAM ranks second among learned estimators (Astrid, DREAM, LPLM) by directly establishing relationships between the query patterns and cardinalities, while LPLM performs the worst, particularly in WIKI, as its wildcard-based decomposition increases prediction difficulty.
Notably, the maximum q-error of SSCard outperforms other estimators, being \textbf{50\%} of the second lowest on TPCH-PN and \textbf{3 $\sim$ 33\%} of the second lowest on other datasets. This improvement is attributed to SSCard's error-bounded estimation. For a a pattern with an actual value $y$, the q-error is bounded by $1+\frac{2\varepsilon\cdot |P|}{y-2\varepsilon\cdot |P|}\leq 1+2\varepsilon\cdot |P|$. For instance, the bound of q-error for SSCard does not exceed $1+2\cdot 32\cdot 8=513$ in DBLP-AN.

\begin{table}[htb]
\centering
\caption{Size of estimators (MB)}
\vspace{-0.3cm}
\label{tab:size}
\resizebox{\columnwidth}{!}{
\begin{tabular}{|c|c|c|c|c|c|}
    \hline
    Estimator & DBLP-AN & IMDB-AN & IMDB-MT & TPCH-PN & WIKI \\
    \hline
    \hlblue{MO} & 7.70 & 13.80 & 24.70 & 15.31 & 365.51 \\
    LBS & 20.85 & 20.69 & 23.04 & 3.56 & 89.95\\
    Astrid & \textbf{0.36} & \textbf{0.36} & \textbf{0.84} & 0.40 & 14.69\\
    DREAM & 7.82 & 7.82 & 7.82 & 7.82 & \textbf{7.88}\\
    LPLM & 1.02 & 1.02 & 1.05 & 1.01 & 9.82 \\
    SSCard & \hlgreen{4.40} & \hlgreen{4.29} & \hlgreen{4.75} & \textbf{0.37} & \hlgreen{23.00}\\
    \hline
\end{tabular}
}
\end{table}

\stitle{Q-error Varying Length of Pattern Strings:} Figure \ref{fig:q_error_l} illustrates the average q-error of each estimator across different pattern lengths. 
SSCard consistently achieves the lowest q-error, particularly for short patterns. This is because SSCard combines both the advantages of the suffix tree and the FM-Index, in which the former provides exact cardinality for query patterns shorter than $h$, 
and the latter supports arbitrary length pattern strings flexibly. Notice that the q-error for learned estimators is significantly high for short pattern lengths because of the infrequency of these short patterns, resulting in inadequate training for this portion of the data.

\stitle{Q-error Varying Actual Result Cardinality:} As shown in Figure~\ref{fig:q_error_c}, SSCard has the lowest q-error especially in low and high cardinalities. 
The results are similar to those of Figure~\ref{fig:q_error_l}, which further verify our previous claim. It is not easy for a learned estimator to produce accuracy results for large cardinalities. Performance degradation may be due to the limited training instances from very large valid strings or the skew data distribution between the training and test set.

\begin{figure*}[t]
    \centering
    \includegraphics[width=\linewidth]{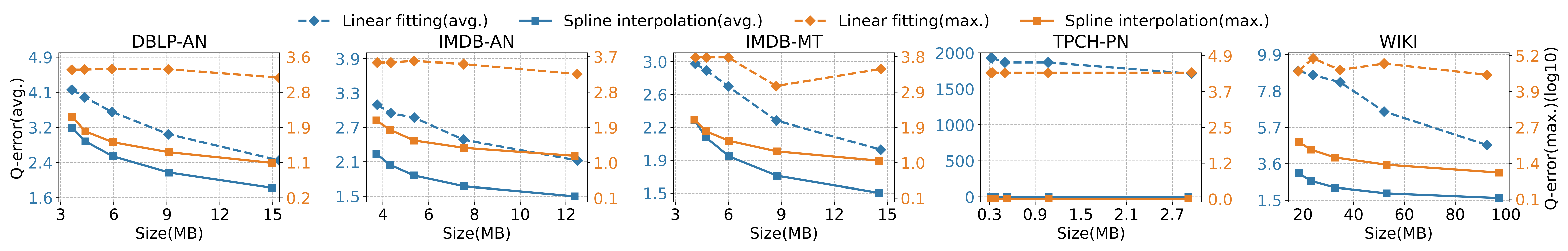}
    \vspace{-0.8cm}
    \caption{Improvement of spline interpolation}
    \label{fig:spline}
\end{figure*}

\begin{figure*}[t]
    \centering
    \includegraphics[width=\linewidth]{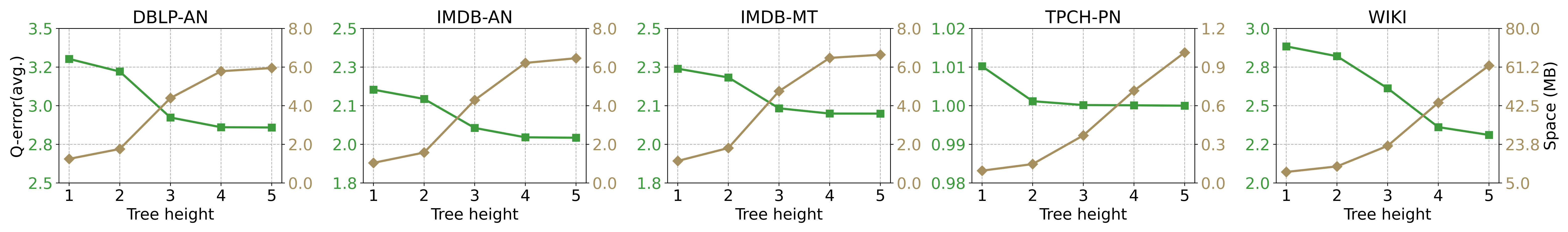}
    \vspace{-0.8cm}
    \caption{\hlblue{S}\hlred{tudy of the effect of the Suffix tree height $h$.}}
    \label{fig:suffix_tree}
\end{figure*}

\stitle{Runtime Analysis:} The runtime of all methods are shown in Table~\ref{tab:runtime}. The pre-processing time for LBS and SSCard refers to the time in building summary data structures, and for other learning-based methods, it refers to the time in generating training data. 
SSCard exhibits a significant advantage in pre-processing and training time except on TPCH-PN datasets. In particular, on DBLP-AN, IMDB-AN and IMDB-MT, SSCard only takes two minutes to construct, achieving an improvement $19\times \sim 40\times$ over the fastest competitior. The training time of SSCard is short because one only needs one pass of the suffix tree to construct spline functions. 
LBS performs well except for WIKI, while large datasets require considerable time to generate $k$-grams. Astrid consumes its major part of the time learing embedding of data strings. For example, Astrid takes 8 hours in learning embedding in WlKl with the top 10\% frequently substrings for training. Both DREAM and LPLM train the model directly on pattern strings, taking the short time in training due to the small size of pattern string set. However, they need to preprocess the original data to produce ground-truth labels for learning. For query time, SSCard and LBS are much faster than the other three learning-based methods. This is because they do not have high computational demands during inference, or vector operations like padding and generating one-hot vectors for pattern strings.

\stitle{Estimator Size:} The space requirements 
are depicted in Table \ref{tab:size}. As analyzed before, the space complexity of SSCard is directly proportional to the size of the data set, the key parts being the fitting functions. It is obvious that the number of fitting functions grows with the increase of the dataset. 
For LBS, longer $k$-grams will occupy more space, as all extended grams and their corresponding hash values need preprocessing to ensure performance. For the DREAM model, its size remains approximately 7.8MB across all four datasets. This is attributed to DREAM's LSTM architecture for pattern string embeddings. As indicated in Table \ref{tab:datasets}, the lengths of the pattern strings in the four data sets exhibit similar distributions. Hence, the size of the DREAM model remains relatively constant and is independent of the underlying data string set $\mathcal{D}$.

\begin{table}[htbp]
  \centering
  \caption{\hlblue{Comparison between FM-Index and SSCard (C++)}}
  \vspace{-0.3cm}
  \resizebox{\columnwidth}{!}{
  \begin{tabular}{|c|c|c|c|c|c|}
    \hline
    Estimator & {DBLP-AN} & {IMDB-AN} & {IMDB-MT} & {TPCH-PN} & {WIKI} \\
    \hline
    \multicolumn{6}{|c|}{Space (MB)} \\
    \hline
    FM-Index & 6.39 & 7.56 & 7.58 & 6.12 & 101.14 \\
    SSCard (C++) & 3.47 & 3.42 & 3.74 & 0.31 & 18.95 \\
    \hline
    \multicolumn{6}{|c|}{Pre-processing Time (s)} \\
    \hline
    FM-Index & 1.57 & 1.95 & 1.80 & 1.36 & 30.56 \\
    SSCard (C++) & 7.10 & 6.32 & 4.57 & 5.35 & 684.77\\
    \hline
    \multicolumn{6}{|c|}{Query Time ($\mu s$)} \\
    \hline
    FM-Index & 1.11 & 1.13 & 1.25 & 0.78 & - \\
    SSCard (C++) & 2.72 & 2.83 & 2.81 & 1.10 & 4.13 \\
    \hline
  \end{tabular}
  }
  \label{tab:FM-Index & SSCard}
\end{table}

\stitle{\hlblue{Comparing SSCard to FM-Index}}\hlblue{: We also compare SSCard with the FM-Index. \texttt{csa\_wt} class of sdsl-lite}~\cite{simpson2010efficient}\hlblue{ in C++ is the state-of-the-art implementation, which is also adopted in the experimental evaluation of}~\cite{loukides2023text}\hlblue{. We test FM-index on string columns in DB by concatenating these strings with special delimiter characters into a long string, as FM-Index does not natively support multiple strings. To ensure a fair comparison, we also implement SSCard in C++. As shown in Table}~\ref{tab:FM-Index & SSCard}, 
\hlblue{while the FM-Index offers faster construction and slightly quicker query times, 
these differences are generally outweighed by SSCard’s significant advantage in space efficiency, particularly for datasets with large or complex alphabets. This is especially evident in WIKI, where the FM-Index 
fails to handle variable-length encodings like UTF-8. For comparison purposes, we encode the WIKI dataset as single-bit characters to enable FM-Index construction, allowing us to evaluate relative space usage.}

\subsection{Effectiveness of Different Components of SSCard
}

\begin{table}[h]
\centering
\caption{\hlblue{Size comparison between \texttt{gzip} and SSCard (MB)}}
\vspace{-0.3cm}
\label{tab:compression}
\setlength{\tabcolsep}{3pt}
\resizebox{\columnwidth}{!}{
\begin{tabular}{|c|c|c|c|c|c|}
\hline
Method & DBLP-AN & IMDB-AN & IMDB-MT & TPCH-PN & WIKI\\
\hline
gzip & 7.13 & 9.61 & 9.80 & 5.24 & 1373.50 \\ \hline
SSCard & 4.40 & 4.29 & 4.75 & 0.37 & 23.00 \\
\hline
\end{tabular}
}
\end{table}

\stitle{Effectiveness of Spline Interpolation:} 
As shown in Figure \ref{fig:spline}, we validate the improvement of the spline interpolation-based strategy over linear fittings. For each dataset, we construct SSCards based on spline interpolation with $\epsilon=4,8,16,32,64$. We also construct a variant of SSCard based on linear fitting, using a comparable space budget by adjusting the number of segments. Spline interpolation yields significantly better performance, primarily because it ensures that fitting errors remain below the given maximum threshold. This prevents sudden error increases and enables SSCard to achieve consistent and controllable results. \hlblue{We also compare the space consumption between SSCard and standard compression technique in python in Table}~\ref{tab:compression}\hlblue{. We take \texttt{gzip} to compress the checkpoints of $Rank(c,i)$ for every $k$ entries as well as the $L$-array. For every query, we decompress the $j=\lfloor i/k \rfloor$-th checkpoint and the corresponding fragment of the $L$-array to accurate calculate $Rank(c,i)$. A smaller $k$ results in faster query time, but with higher space consumption. We set $k=500$ to achieve comparable query time between SSCard and \texttt{gzip}. Under this configuration, the gzip-based approach consumes significantly more space than SSCard.}

\stitle{\hlblue{S}\hlred{uffix Tree height $h$:}}\hlred{ As shown in Figure}~\ref{fig:suffix_tree}\hlred{, we evaluate SSCard under different values of the suffix tree height $h$. As $h$ increases, the q-error decreases, but the space usage grows accordingly. This is because more queries can be estimated precisely on the pruned suffix tree with larger height. When $h=3$, SSCard can estimate most of the short strings accurately, avoiding the impact of multiple occurrences of $P$ in a single data string. However, the benefit of increasing $h$ diminishes beyond this point, while the space overhead continues to grow. 
Therefore, we set $h=3$ for our evaluation.}

\begin{table}[h]
\centering
\caption{\hlblue{C}\hlyellow{omparison between SSCard w/o pushup operation}}
\vspace{-0.3cm}
\label{tab:pushup}
\setlength{\tabcolsep}{3pt}
\resizebox{\columnwidth}{!}{
\begin{tabular}{|c|c|c|c|c|c|}
\hline
Method & DBLP-AN & IMDB-AN & IMDB-MT & TPCH-PN & WIKI\\
\hline
\multicolumn{6}{|c|}{Space (MB)} \\
\hline
w/o-pushup & 8.46 & 8.95 & 13.75 & 0.99 & 346.46 \\ 
pushup & \hlgreen{4.40} & \hlgreen{4.29} & \hlgreen{4.75} & 0.37 & \hlgreen{23.00} \\
\hline
\multicolumn{6}{|c|}{Q-error (Avg.)} \\
\hline
w/o-pushup & 2.72 & 1.92 & 2.01 & 1.00 & 2.44\\
pushup & 2.92 & 2.05 & \hlgreen{2.14} & 1.00 & 2.61\\
\hline
\end{tabular}
}
\end{table}

\stitle{Effect of Pushup:} As mentioned in Section \ref{subsec:build_sscard}, SSCard takes a pushup strategy to put rare characters into the segments linking from the high-level tree nodes adaptively. 
\hlblue{T}\hlyellow{able}~\ref{tab:pushup}\hlyellow{ illustrates an obvious reduction in space by the pushup operation, with minor increase in q-error.} For example, the optimized space cost is reduced by 91\% on WIKI compared to the space cost without pushup, which requires storing $O(|\Sigma|)$ extra space cost per segment, which becomes substantial when the alphabet is large.

\begin{figure}[htbp]
  \centering

  \hspace{-0.03\linewidth}
    \begin{subfigure}[b]{0.45\linewidth}
    \centering
    \includegraphics[width=\linewidth]{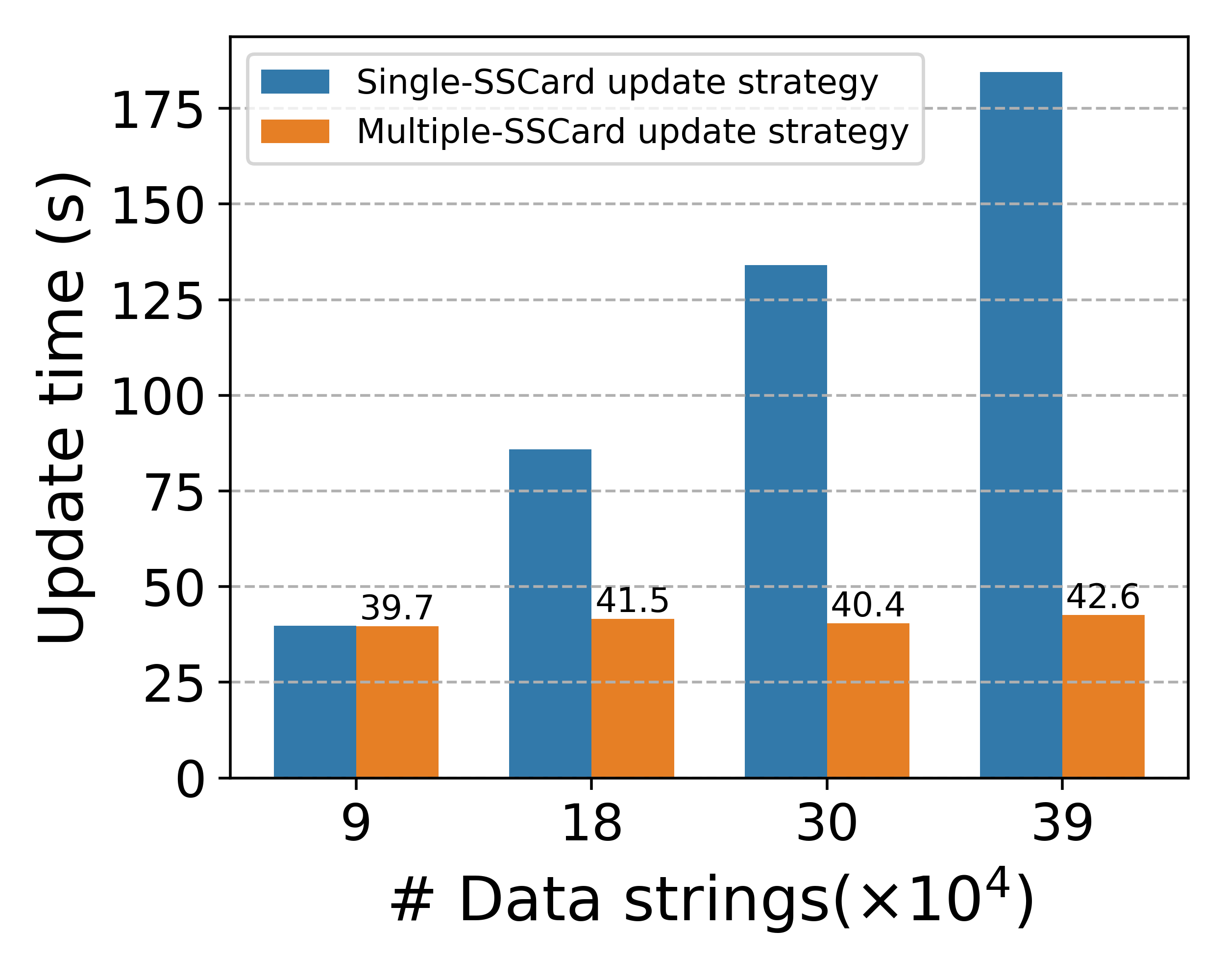}
    \vspace{-0.7cm}
    \caption{Incremental update time}
    \label{subfig:inc_r_time}
    \end{subfigure}
    \hspace{0.04\linewidth}
    \begin{subfigure}[b]{0.45\linewidth}
    \centering
    \includegraphics[width=\linewidth]{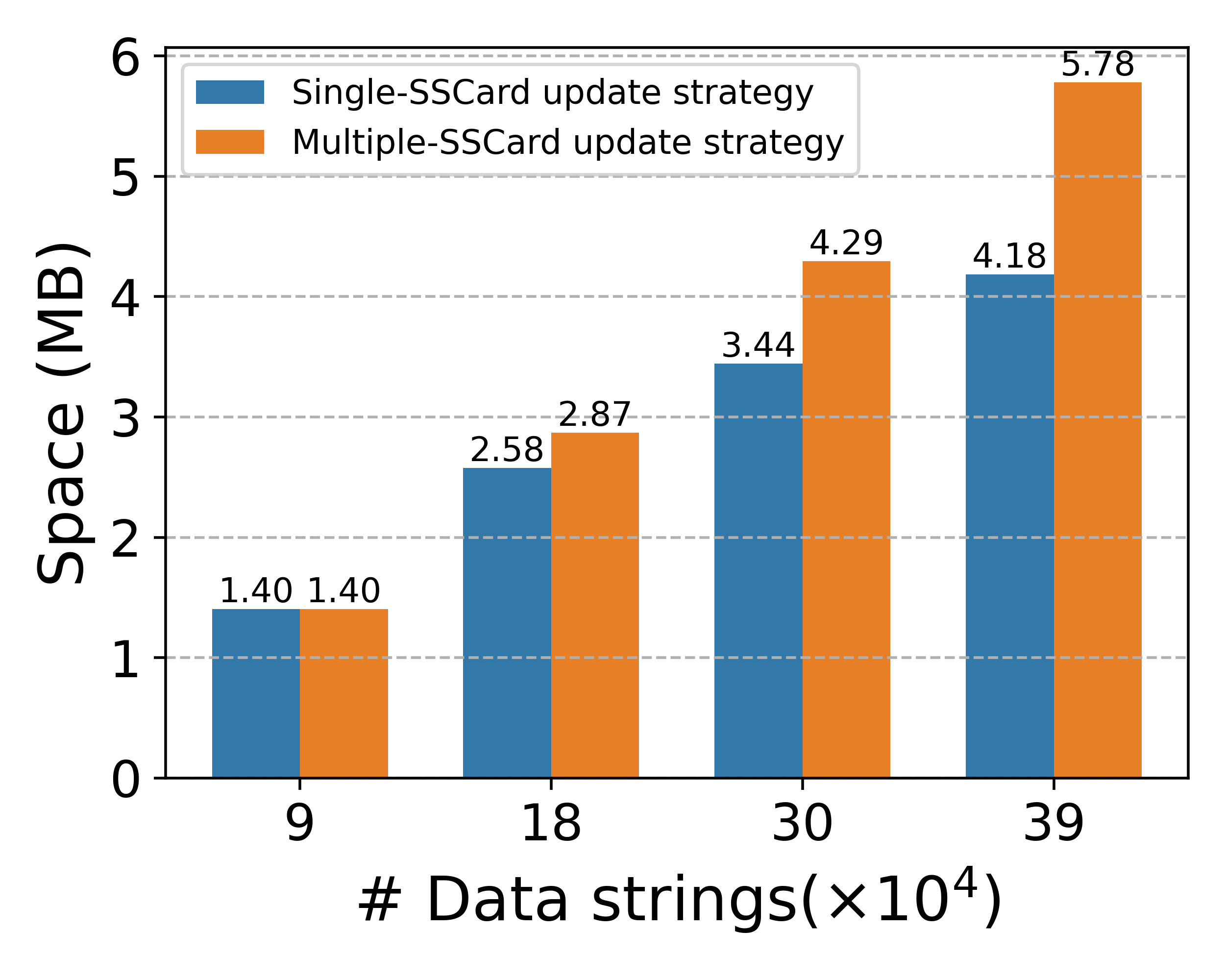}
    \vspace{-0.7cm}
    \caption{Space}
    \label{subfig:inc_space}
    \end{subfigure}
  \hfill
    \begin{subfigure}[b]{0.45\linewidth}
    \centering
    \includegraphics[width=\linewidth]{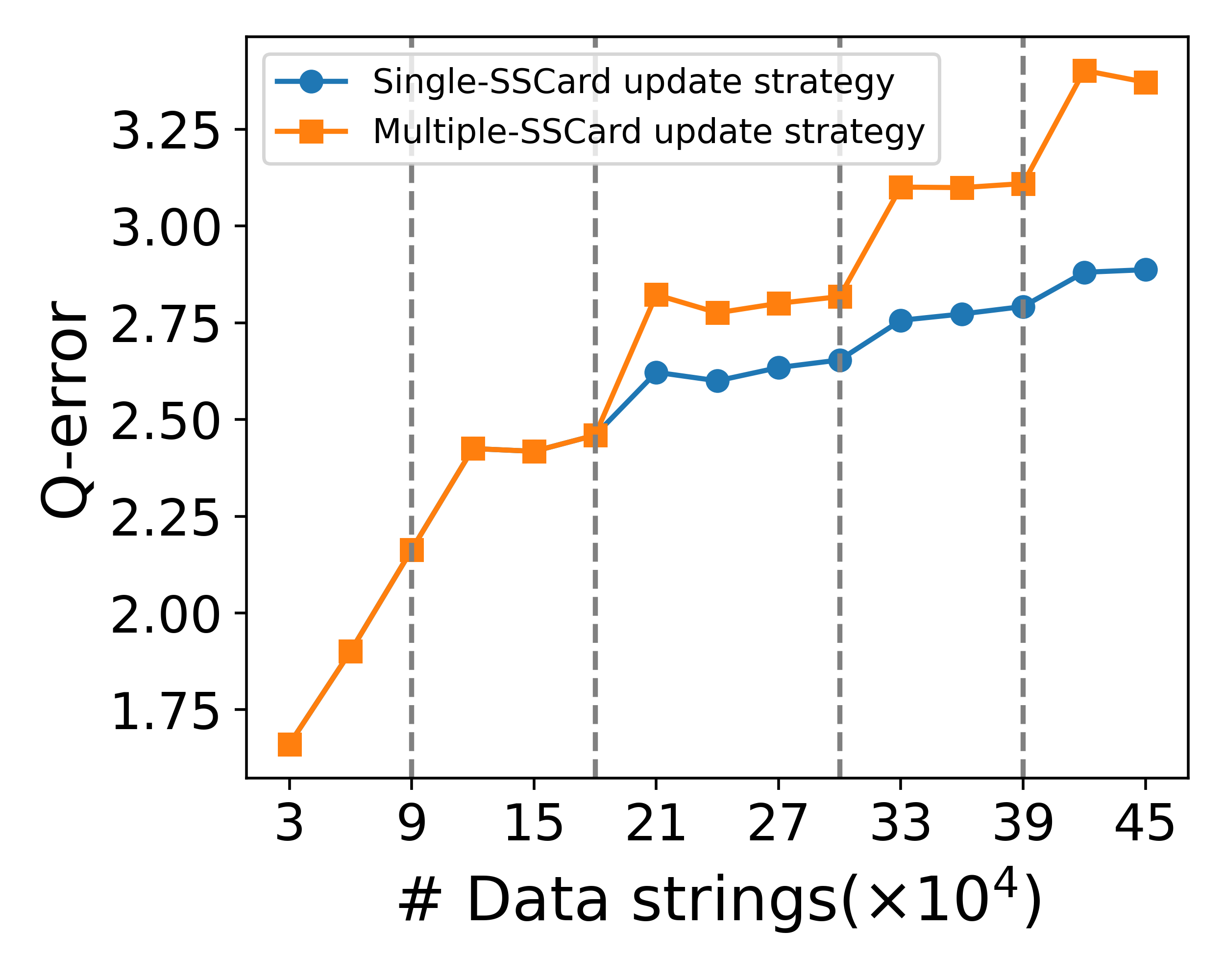}
    \vspace{-0.7cm}
    \caption{Q-error}
    \label{subfig:inc_q_error}
  \end{subfigure}
  \hfill
  \begin{subfigure}[b]{0.45\linewidth}
    \centering
        \includegraphics[width=\linewidth]{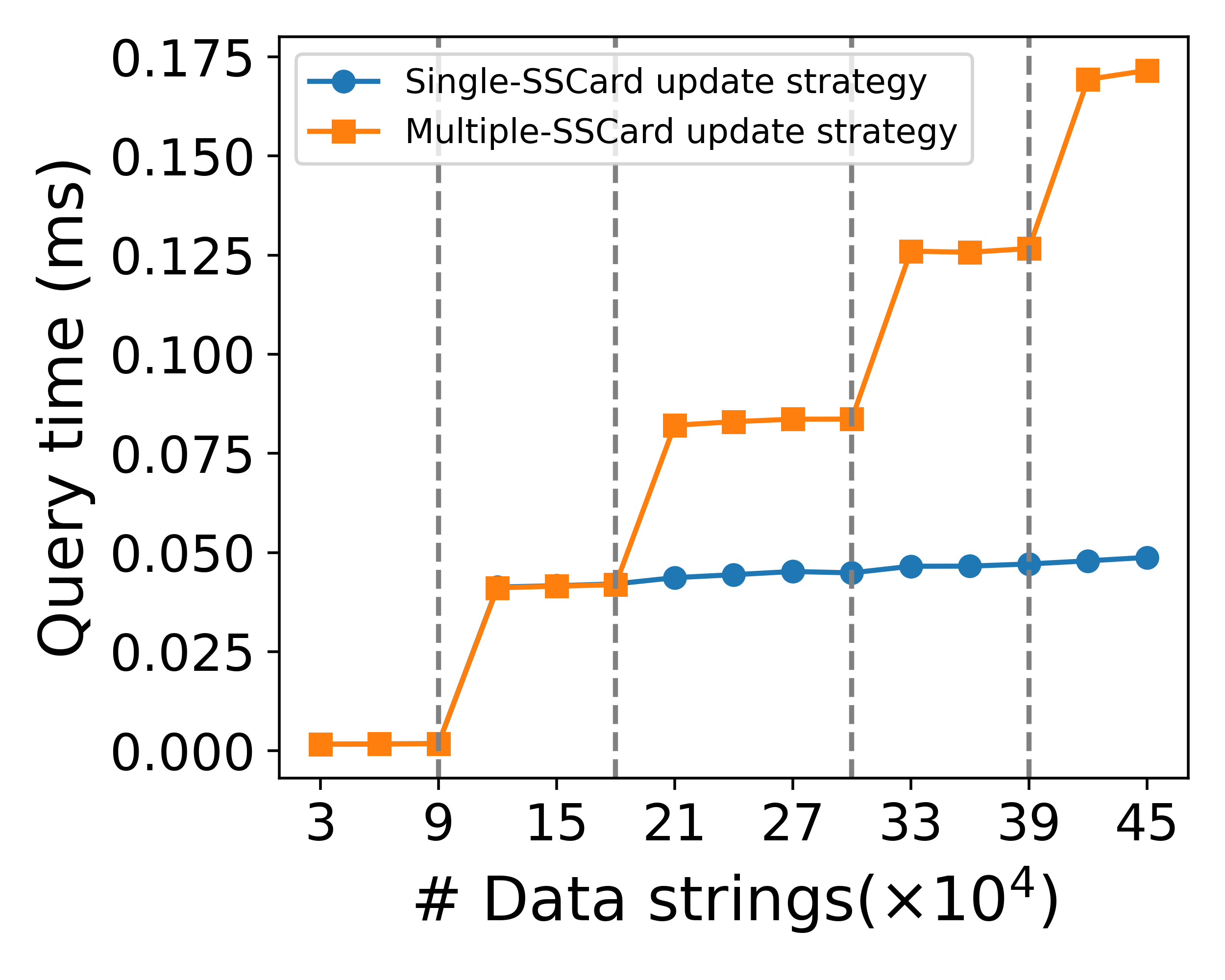}
    \vspace{-0.7cm}
    \caption{Query time}
    \label{subfig:inc_q_time}
  \end{subfigure}
  \hfill
\vspace{-0.3cm}
  \caption{\hlblue{I}\hlyellow{n}\hlred{cremental update (DBLP-AN)}}
  \label{fig:inc}
\end{figure}

\stitle{\hlblue{E}\hlyellow{v}\hlred{aluation on Incremental Update:}}\hlred{ We mainly compare two update strategies, including the single-SScard update strategy and multiple-SSCard update strategy, in terms of the incremental update cost, space usage, q-error and query time. The experiments are conducted on the DBLP-AN dataset. We divide data strings into 15 chunks, each of which fits into main memory. 
These chunks are then fed into the estimator sequentially. 
When the number of nodes in the pruned suffix tree $T_i$ exceeds $2.5 \times 10^5$, we either rebuild the existing SSCard or create a new one, depending on the update strategy.
As shown in Figure}~\ref{fig:inc}\hlred{, although the multiple-SSCard update strategy incurs slightly higher q-error (Figure}~\ref{subfig:inc_q_error}\hlred{) and slower query time (Figure}~\ref{subfig:inc_q_time}\hlred{) compared to the single-SSCard update strategy, it achieves much better update efficiency. Notably, the update time remains nearly constant as the data grow, while the single-SSCard update time increases linearly (Figure}~\ref{subfig:inc_r_time}\hlred{). 
Overall, despite minor trade-offs, the multiple-SSCard update strategy is several times faster in updates and thus more practical for dynamic scenarios, especially given that update costs dominate over query time.}

\subsection{End-to-End Query execution Time}
We evaluate the end-to-end query execution time by injecting estimated cardinalities into PostgreSQL 14.5. We take IMDB as dataset and select 79 out of 116 queries of the JOB workload~\cite{goodplan2015} that contains \texttt{LIKE} statement with the form $\%word\%$. \hlyellow{Cardinalities for these \texttt{LIKE} predicates are estimated by SSCard as well as other baselines, and then injected into query plans for execution.} As shown in Table \ref{tab:end-to-end}, the first row presents the original cardinalities estimated by PostgreSQL. \hlyellow{In the following rows, the cardinalities of the \texttt{LIKE} predicates are estimated using different methods, while the cardinalities of the other filters are still taken from PostgreSQL's estimates. Note that the last row represents the true cardinalities of \texttt{LIKE} predicates.} To ensure that the cardinalities injected to single tables are not affected by other predicates, we modify the queries so that tables containing the \texttt{LIKE} predicate do not include any other predicates. We run each method three times and analyze the average execution time. The injection approach is the same as~\cite{aytimur2024lplm}. Compared to ~\cite{aytimur2024lplm}, we construct SSCard on all columns that filtered with \texttt{LIKE}. 

As illustrated in Table~\ref{subtab:end-to-end-query}, \hlyellow{SSCard+PG improved most of the queries, but the execution time is not strictly correlated with the estimation accuracy of \texttt{LIKE} prediates (Table}~\ref{subtab:end-to-end-time}\hlyellow{). 
This may be because PostgreSQL produces inaccurate join cardinality estimates, even when provided with more accurate or true single-table cardinalities.} 
A typical example is 4a.sql, PG underestimates the join results of table \texttt{info\_type} and \texttt{movie\_info\_idx}, and after injecting true cardinalties larger than PG's of \texttt{keyword.keyword LIKE '\%sequel\%'}, PG put this filter in a later section of the query plan and thus produces numerous intermediate results. This also shows that a more effective benchmark for single-table cardinality estimation is needed.

\begin{table}[htb]
\caption{\hlyellow{Experiments on the Join Order Benchmark}}
\vspace{-0.3cm}
\label{tab:end-to-end}
\begin{subtable}[t]{0.45\textwidth}
    \centering
    \resizebox{\columnwidth}{!}{
        \begin{tabular}{|c|c|c|c|c|}
            \hline
            \multirow{2}{*}{} & \multicolumn{2}{|c|}{Improved} & \multicolumn{2}{|c|}{Regressed} \\
            \cline{2-5}
            & \#Queries & Avg. Imp. & \#Queries & Avg. Reg. \\
            \hline
            PostgreSQL & - & - & - & -\\
            \hline
            MO+PG & 44 & 9.29\% & 35 & 117.03\%\\
            \hline
            LBS+PG & 43 & 9.10\% & 36 & 68.2\%\\
            \hline
            Astrid+PG & 35 & 20.36\% & 44 & 90.05\%\\
            \hline
            DREAM+PG & 43 & 12.77\% & 36 & 26.61\%\\
            \hline
            \hlgreen{SSCard+PG} & 50 & 4.91\% & 29 & 85.29\% \\
            \hline
            TrueCard+PG & 50 & 8.50\% & 29 & 74.40\% \\
            \hline
        \end{tabular}
    }
    \vspace{0.cm}
    \caption{Improved and regressed queries}
    \label{subtab:end-to-end-query}
\end{subtable}

\begin{subtable}[t]{0.45\textwidth}
    \centering
    \resizebox{\columnwidth}{!}{
        \begin{tabular}{|c|c|c|c|c|c|c|}
            \hline
            & Avg. & 50th & 90th & 99th & Max. & Total\\
            \hline
            PostgreSQL & 1.34 & 0.34 & 4.43 & 8.82 & 11.00 & 105.83\\
            \hline
            MO+PG & 1.45 & \textbf{0.32} & 4.68 & 9.80 & \textbf{10.58} & 114.64\\
            \hline
            LBS+PG & 1.37 & 0.35 & \textbf{4.32} & 8.75 & 11.33 & 108.52\\
            \hline
            Astrid+PG & 1.48 & 0.33 & 5.06 & 10.15 & 10.88 & 116.58 \\
            \hline
            DREAM+PG & \textbf{1.28} & 0.33 & \textbf{4.32} & 8.87 & 11.00 & \textbf{101.20}\\
            \hline
            \hlgreen{SSCard+PG} & 1.35 & 0.36 & 4.39 & \textbf{8.67} & 10.80 & 106.66 \\
            \hline   
            TrueCard+PG & 1.37 & 0.38 & 4.40 & 8.78 & 10.98 & 108.29 \\
            \hline
        \end{tabular}
    }
    \vspace{0.cm}
    \caption{Query execution time}
    \label{subtab:end-to-end-time}
\end{subtable}

\end{table}

\section{Related Work}
\label{sec:related}

Here, we review the progress in the string index, the traditional and learned string cardinality estimation, and the learned index.

\stitle{String Index.} Various forms of string indexes and their optimization strategies, such as suffix tree, $k$-gram, and BWT ~\cite{burrows1994block} have been extensively studied. 
Ukkonen algorithm builds the suffix tree by extending character iteratively with $O(m)$ time and space cost, where $m$ is the string length. Suffix arrays offer a space-efficient structure with $O(m)$ space. 
BD-anchors ~\cite{Loukides2021} reduces the space consumption in $k$ gram algorithms, and were further optimized to reduce construction cost~\cite{loukides2023text} recently. However, BD-anchors require that the pattern string should be longer than a given threshold. FM-Index~\cite{ferragina2000opportunistic} with the extra cached cumulative ranked on the $L$ array to achieve the $O(1)$ rank operation, whose space is not neglected when $\Sigma$ is large. Wavelet tree~\cite{grossi2003high} achieves $O(|\mathcal{D}|)$ extra space with $O(\log(|\Sigma|)$ rank operation. 

\stitle{Traditional Cardinality Estimation for String.} 
String cardinality estimation applies string indexing techniques in DBMS. KVI~\cite{krishnan1996estimating}, MO~\cite{jagadish1999substring}, and CRT~\cite{gravanoselectivity} mainly use the suffix tree to estimate pattern cardinality, differing in how they compute dependency probabilities when pattern length exceeds the tree height. SPH~\cite{aytimur2018estimating} and \hlyellow{P-SPH}~\cite{aytimur2021using} \hlyellow{use} $k$-grams to discover frequent patterns in the data strings and cache the cardinality of these patterns in histograms, but face challenges with both high mining costs and long query patterns. 
LBS [24] combines $k$-grams and set hashing for longer queries. 
We can see that considerable effort has been devoted to long patterns in classic methods. 
Luckily, SSCard extends FM-Index, which naturally supports various-length patterns.

\stitle{Learned Cardinality Estimation for String.} Astrid~\cite{shetiya2020astrid} is one of the first deep-learning works for string cardinality estimation. It learns the selectivity-aware embeddings for the substrings, following a regression network to predict substring cardinality. CardNet~\cite{wang2020monotonic} takes an encoder-decoder structure 
to build the relationships between pattern strings and their cardinalities. DREAM~\cite{kwon2022cardinality} shares a similar idea as CardNet, but uses the sequence model (LSTM) in the encoder-decoder framework. 
The most recent work for LIKE predicates, LPLM~\cite{aytimur2024lplm} defines a probability distribution function to capture the semantics of LIKE-patterns. 

\stitle{Learned Index.} The learned indexes attempt to predict the position of the tuple using a learned model with a tuple search key as input. 
The early work is based on the neural network~\cite{learnedindex2018}, which verifies the feasibility of the learned index. To achieve efficient construction and provide error bounds, methods like FITing-Tree~\cite{galakatos2019fiting} and PGM-Index~\cite{ferragina2020pgm} approximate CDF (cumulative distribution function) with linear functions, while spline interpolation has also proven effective. Leveraging the similarity between the rank order of the characters in the $L$-array and the CDF, 
we adopt greedy spline interpolation \cite{Spline} to balance space consumption and estimation quality. \hlblue{It is worth noting that LISA}~\cite{ho2019lisa} \hlblue{is a learned index also based on the FM-Index. It achieves $O(|Q|/K)$ time complexity for a string $Q$ by searching a $K$-length substring each time and applying RMI}~\cite{kraska2018case}\hlblue{ to locate the index of the substring in the $L$-array. 
However, LISA takes larger space than the FM-index and faces the similar challenge in handling large alphabets.}

\section{Conclusion and Future Work}
\label{sec:conclude}

We proposed SSCard, a cardinality estimator that optimizes FM-Index by incorporating a pruned suffix tree and a learned representation. We also introduced a bidirectional estimation algorithm and incremental update strategies. 
SSCard can be further investigated in the following directions. First, SScard is expected to support multiple patterns with wildcards by extending the $L$-triples with string IDs and pattern indexes. 
Second, 
we plan to utilize SSCard to estimate the join cardinality of two tables with string columns, as well as the single table cardinality filtered with multi predicates on different string columns. 

\section{Acknowledgement}
\label{sec:ack}
This work was supported in part by NSFC under Grant No.62272008 and CCF-Huawei Populus Grove Fund.

\bibliographystyle{ACM-Reference-Format}
\bibliography{BibTeX}


\end{document}